\documentclass[]{aa}

\usepackage[]{graphicx}  

\usepackage{natbib,twoopt}
\bibpunct{(}{)}{;}{a}{}{,} 

\usepackage[varg]{txfonts}
\usepackage{color,linegoal}  

\usepackage[normalem]{ulem}  

\def\pow#1#2{#1$\times$10$^{#2}$}
\def\powm#1#2{#1\times10^{#2}}
\def\micron{$\mu$m}
\def\kms{$\mathrm{km}\,\mathrm{s}^{-1}$}  
\def\psqcm{$\mathrm{cm}^{-2}$}
\def\pccm{$\mathrm{cm}^{-3}$}
\def\Tdust{$T_\mathrm{dust}$}  
\def\Tgas{$T_\mathrm{gas}$} 

\def\vlsr{$V_\mathrm{lsr}$}  
\def\taudv{$\int \tau_\nu \,\mathrm{d}V$}  
\def\Wsqm{W\,m$^{-2}$}  
\def\h{$^\mathrm{h}$}  
\def\m{$^\mathrm{m}$}  
\def\Msun{$M_\odot$}  

\definecolor{orange}{rgb}{0.8,0.4,0.0}
\definecolor{darkblue}{rgb}{0.0,0.0,0.6}
\definecolor{darkred}{rgb}{0.75,0.0,0.0}

\def\Hii{H{\sc ii}}
\def\HH{H$_2$}
\def\HCOplus{HCO$^+$}

\def\thCO{$^{13}$CO}

\def\water{H$_2$O}
\def\waterp{H$_2$O$^+$}
\def\OHp{OH$^+$}
\def\CHplus{CH$^+$}
\def\Catom{C{\sc i}}
\def\COtwo{CO$_2$}

\def\markchanges{yes}  
\def\marked{yes}
\def\unmarked{no}
\ifx\markchanges\marked 
	\newcommand{\removed}[1]{\textcolor{darkred}{[}\sout{#1}\textcolor{darkred}{]}}  
	
\fi
\ifx\markchanges\unmarked 
	\newcommand{\removed}[1]{}  
	
\fi

\begin{document}

\title{Three-dimensional distribution of hydrogen fluoride gas toward NGC\,6334\,I and I(N)\thanks{{\it Herschel} is an ESA space observatory with science instruments provided by European-led Principal Investigator consortia and with important participation from NASA.}}
\titlerunning{Three-dimensional distribution of HF gas toward NGC~6334}

\author{
M.~H.~D.~van der Wiel\inst{\ref{nbi-starplan},\ref{lethbridge}} 
\and
D.~A.~Naylor\inst{\ref{lethbridge}}
\and
G.~Makiwa\inst{\ref{lethbridge}}
\and
M.~Satta\inst{\ref{rome}}
\and
A.~Abergel\inst{\ref{ias}}
}

\institute{
Centre for Star and Planet Formation, Niels Bohr Institute \& Natural History Museum of Denmark, University of Copenhagen, \O ster Voldgade 5--7, DK-1350 \mbox{Copenhagen~K}, Denmark
\label{nbi-starplan}
\and
Institute for Space Imaging Science, Department of Physics \& Astronomy, University of Lethbridge, Canada
\label{lethbridge}
\and 
CNR-ISMN, Department of Chemistry, The University of Rome ``Sapienza,'' P.le A.~Moro 5, I-00185 Rome, Italy
\label{rome}
\and 
Institut d'Astrophysique Spatiale, CNRS, Univ.~Paris-Sud, Universit\'e Paris-Saclay, B\^at.~121, 91405, Orsay Cedex, France
\label{ias}
}

\date{\today}

\abstract
{ 
The HF molecule has been proposed as a sensitive tracer of diffuse interstellar gas, while at higher densities its abundance could be influenced heavily by freeze-out onto dust grains. 
}
{
We investigate the spatial distribution of a collection of absorbing gas clouds, some associated with the dense, massive star-forming core NGC\,6334\,I, and others with diffuse foreground clouds elsewhere along the line of sight. For the former category, we aim to study the dynamical properties of the clouds in order to assess their potential to feed the accreting protostellar cores.
}
{
We use far-infrared spectral imaging from the \mbox{\it Herschel} SPIRE iFTS to construct a map of HF absorption at 243 \micron\ in a 6\arcmin$\times$3\farcm5  region surrounding NGC\,6334~I and I(N). 
}
{
The combination of new, spatially fully sampled, but spectrally unresolved mapping with a previous, single-pointing, spectrally resolved HF signature yields a three-dimensional picture of absorbing gas clouds in the direction of NGC\,6334. 
Toward core~I, the HF equivalent width matches that of the spectrally resolved observation. At angular separations $\gtrsim$20\arcsec\ from core I, the HF absorption becomes weaker, consistent with three of the seven components being associated with this dense star-forming envelope. Of the remaining four components, two disappear beyond $\sim$1\arcmin\ distance from the NGC\,6334 filament, suggesting that these clouds are spatially associated with the star-forming complex. 
Our data also implies a lack of gas phase HF in the envelope of core I(N). Using a simple description of adsorption onto and desorption from dust grain surfaces, we show that the overall lower temperature of the envelope of source I(N) is consistent with freeze-out of HF, while it remains in the gas phase in source I. 
}
{
We use the HF molecule as a tracer of column density in diffuse gas ($n_\mathrm{H}$$\approx$$10^2$--$10^3$~\pccm), and find that it may uniquely trace a relatively low density portion of the gas reservoir available for star formation that otherwise escapes detection. At higher densities prevailing in protostellar envelopes ($\gtrsim$$10^4$~\pccm), we find evidence of HF depletion from the gas phase under sufficiently cold conditions. 
}


\maketitle

\section{Introduction}
\label{sec:intro} 

The hydrogen fluoride molecule, HF, was first observed in the interstellar medium by \citet{neufeld1997b} with the {\it Infrared Space Observatory} \citep[{\it ISO},][]{kessler1996}. While {\it ISO} had a wavelength range that encompassed only the $J$=2--1 rotational transition of HF, the next observatory able to observe HF -- the {\it Herschel} Space Observatory \citep{pilbratt2010} -- covered longer far-infrared wavelengths, and it thus opened up access to the ground-state rotational transition, $J$=1--0, at 1232.48~GHz (243.24~\micron). {\it Herschel} has observed HF in absorption along many lines of sight, both inside the Galaxy  \citep{neufeld2010b,sonnentrucker2010,sonnentrucker2015,philips2010,kirk2010,monje2011a,emprechtinger2012,lopez-sepulcre2013a,goicoechea2013} and in nearby extragalactic objects \citep{rangwala2011,kamenetzky2012,rosenberg2014a,monje2014}. HF absorption has even been detected with ground-based observatories: \citet{monje2011c} have made use of the substantial redshift of the Cloverleaf quasar at $z$=2.56 shifting the HF~1--0 line into the submillimeter window attainable with the CSO on Mauna Kea, and \citet{kawaguchi2016} detect it in the $z$=0.89 absorber toward PKS1830$-$211, using ALMA in the Chilean Atacama desert. 
Because of its large dipole moment and high Einstein $A$ coefficient for radiative decay, rotational states $J$$\neq$0 of HF only become significantly populated in highly energetic conditions. It is for this reason that HF has been clearly detected in emission in a mere handful of cases: in the inner region of an AGB star's envelope \citep[IRC+10216,][]{agundez2011}, in the Orion Bar photodissociation region \citep{vandertak2012a}, and in an external galaxy harboring an actively accreting black hole \citep[Mrk\,231,][]{vanderwerf2010}. 

Atomic fluoride, F, has a unique place in the interstellar chemistry of simple molecules. It is the only element which, simultaneously,   
(1) is mainly neutral because of its ionization potential $>$13.6~eV, 
(2) reacts exothermically with \HH\ -- unlike \emph{any} other neutral atom -- to form its neutral diatomic hydride HF, and 
(3) lacks an efficient chemical pathway to produce its hydride cation HF$^+$ due to the strongly endothermic nature of the reaction with H${_3}^+$. We refer to \citet{neufeld2009b}, references therein, and the comprehensive review by \citet{gerin2016} for more details on the chemistry of HF and a comparison with other hydride molecules. For the reasons listed above, chemical models predict that essentially all interstellar F is locked in HF molecules \citep{zhu2002,neufeld2005}, which has been confirmed by observations across a wide range of atomic and molecular ISM conditions \citep[e.g.,][]{sonnentrucker2010,sonnentrucker2015}. With recent experimental results by \citet{tizniti2014} showing that, especially at low temperatures approaching 10~K, the reaction F + \HH\ $\rightarrow$ HF + H proceeds somewhat slower than earlier assumptions, chemical models are now able to reproduce HF/\HH\ ratios of $\sim$\pow{1}{-8}, measured most directly by \citet{indriolo2013a}, and observed to be rather stable across different sightlines.  
Interferometric observations show that CF$^+$, the next most abundant F-bearing species after HF, has an abundance roughly two orders of magnitude lower than HF, both inside our Galaxy \citep{liszt2015b} and in an extragalactic absorber \citep{muller2016}. 
As for destruction of HF, the most efficient processes are UV photodissociation and reactions with C$^+$, but both of these are unable to drive the majority of fluoride out of HF, due to shielding, already at modest depths of $A_V>0.1$ \citep{neufeld2005}. 


Because of the constant HF/\HH\ abundance ratio and the high probability that HF molecules are in the rotational ground state, measurements of HF $J$=0$\rightarrow$1 absorption provide a straightforward proxy of \HH\ column density. This has led to the suggestion that, at least in diffuse gas, HF absorption is a more reliable tracer of total gas column density than the widely used carbon monoxide (CO) rotational \emph{emission} lines, and is more sensitive than CH or \water\ absorption \citep[e.g.,][]{gerin2016}. 
Apart from the uncertain and variable CO abundance, local excitation conditions have a profound effect on the level populations of CO, complicating the conversion from observed line strength of a particular CO transition to \HH\ column density \citep{bolatto2013}. The greatest gas-phase CO abundance variations occur in dense, cold regions where CO freezes out onto surfaces of dust grains, proven by observed CO abundances decreasing in the gas phase and increasing in the ice phase as conditions get colder \citep[e.g.,][]{jorgensen2005a,pontoppidan2005a}.  
In addition, the particular fraction of the neutral ISM that is in the diffuse/translucent phase is inconspicuous in CO \citep{bolatto2013}, but is detectable using hydride absorption lines.  

Of course, for absorption line studies, one relies on lines of sight with sufficiently strong continuum background, for example those toward dense star-forming clouds. Such restrictions do not apply for emission line tracers. Besides CO rotational lines, fine structure line emission due to atomic C and the C$^+$ and N$^+$ ions has been used as a tracer of (diffuse) gas throughout the Galaxy \citep[e.g.,][]{langer2014a,velusamy2014b,gerin2015a,goicoechea2015b,goldsmith2015}. For all these tracers, however, the conversion to \HH\ column density depends strongly on physical properties such as ionization fraction and excitation conditions. 

Based on the above arguments, HF absorption measurements are a good tracer of overall gas column density. However, as addressed for example by \citet{philips2010} and \citet{emprechtinger2012}, HF itself may suffer from freeze-out effects as occurs with other interstellar molecules. While studies have been done on the interaction of \water\ with HF as a polluting agent in the Earth's atmosphere \citep{girardet2001}, the density and temperature conditions needed for HF adsorption onto dust grains have not been studied in astrophysical contexts so far. Any freeze-out of interstellar HF will obfuscate the direct connection between HF absorption depth and \HH\ column density described above. 

The well-known progression of pre- and protostellar stages for stars with masses similar to the Sun \citep{shu1977} is not applicable for high-mass stars ($\gtrsim8$~\Msun). In the latter category, protostellar hydrogen fusion starts while accretion from the surrounding gas envelope is still ongoing \citep{palla1993}. In the `competitive accretion' scenario, multiple massive protostars in a clustered environment are fed from the same gas reservoir \citep{bonnell2001b}. For high-mass protostars, material can continuously be added to the gravitationally bound circumstellar envelope which provides the reservoir for further accretion onto the protostar. It is therefore important, particularly for regions of \emph{high-mass} star formation, to study not just the gravitationally bound circumstellar envelopes, but also the dynamical properties of surrounding gas clouds. Especially for the latter component, simple hydride molecules have the potential to reveal gas reservoirs to which emission lines of `traditional' tracer species, such as CO, \HCOplus, and CS, are insensitive due to their relatively high critical densities. In this paper, we investigate two envelopes of (clusters of) protostars embedded in the \object{NGC\,6334} molecular cloud as well as lower density clouds surrounding the dense complex. 

The filamentary, star-forming cloud NGC~6334, at a distance of 1.35~kpc \citep{wu2014}, harbors a string of dense cores, identified in the far-infrared by roman numerals I through VI  \citep{mcbreen1979}, with an additional source identified $\sim$2\arcmin\ north of source I, later named `I(N)' \citep{gezari1982}. 
The larger scale NGC\,6334 filament has an \HH\ column density of $>$\pow{2}{22}~\psqcm\ even at positions away from the embedded cores \citep{russeil2013}.  \citet{zernickel2013} observed the velocity structure of NGC\,6334 at 0.15~pc resolution. These authors explain the velocity profile along  the filament with a cylindrical model collapsing along its longest axis under the influence of gravity. 
In this paper we study specifically the region of $\sim$2.4$\times$1.6~pc surrounding the embedded cores \object{NGC\,6334~I} and \object{NGC\,6334~I(N)}. Source I is host to an ultra-compact \Hii\ region, designated source `F' in a 6~cm radio image of the cloud \citep{rodriguez1982}. Based on multi-wavelength dust continuum measurements, studies by \citet{sandell2000} and \citet{vandertak2013a} have independently determined that the mass of source I(N) exceeds that of sister source I by a factor of $\sim$2--5, but the ratio of their bolometric luminosities is 30--140 in favor of source I, due to the markedly lower temperature for source I(N). As expected for a warm (up to $\sim$100~K), dense, massive star-forming core, NGC\,6334~I is extremely rich in molecular lines, spectacularly demonstrated by the 4300 lines detected in the 480--1907~GHz spectral survey by \citet{zernickel2012}\footnote{See the introduction section of this reference for a list of earlier spectral survey work on NGC\,6334~I.}. 
The differences between the two neighboring cores all suggest that core I is in a more evolved stage of star formation than core I(N). 
Both cores have been studied with radio and (sub)millimeter interferometer observatories, showing that each separates into several subcores at arcsecond resolution \citep[i.e., at scales $\lesssim$0.01~pc,][]{hunter2006,hunter2014,brogan2009}. 

To probe gas clouds in front of the NGC\,6334 complex, absorption measurements have been obtained in lines of several hydrides. 
Spatially extended OH hyperfine line absorption at \mbox{1.6--1.7~GHz} was observed toward the NGC\,6334 filament by \citet{brooks2001}. The spectrally resolved mapping observations from the Australia Telescope Compact Array allowed these authors to ascribe particular velocity components of the absorption to a foreground cloud close to NGC\,6334 and other components to clouds with even larger angular extent. \citet{vanderwiel2010} used the Heterodyne Instrument for the Far-Infrared \citep[HIFI,][]{degraauw2010} onboard {\it Herschel} to study the spectral profile of the rotational ground state lines of CH at 532 and 537 GHz, and found four distinct absorption components overlapping with the velocity range of OH absorption, and one single emission component emanating in core NGC\,6334~I itself. At 1232.5~GHz in the same spectral survey, \citet{emprechtinger2012} find the exact same four absorbing clouds in the HF rotational ground state, and invoke three velocity components to explain the hot core component. While the hot core component(s) appear in emission in CH, they are in absorption in HF, because CH 1$\rightarrow$0 has a lower Einstein $A$ coefficient than HF 1$\rightarrow$0 (see above).

\begin{table*}
\caption{SPIRE spectrometer observations used in this paper.}
\label{t:obs} 
\centering 
\begin{tabular}{l l l l @{\ \ } l l l r}
\hline\hline
		 	&		 	& \multicolumn{3}{c}{Coordinates of central pointing\tablefootmark{a}}	& 		&  \\
\cline{3-5}
{\it Herschel}	& Observation	& Target name & Right Ascension & Declination	& Jiggle	& Gain & \multicolumn{1}{c}{Duration} \\
observation ID	& date		&			& (J2000)		     & (J2000)		& pattern\tablefootmark{b} & mode & \multicolumn{1}{c}{(s)} \\
\hline
1342214827	& 2011-02-26	& NGC\,6334 I	&  17\h20\m54\fs15 & $-$35\degr47\arcmin07\farcs4		& 4$\times$4 & Nominal & 11491	\\
1342214841 	& 2011-02-27	& NGC\,6334 I(N) & 17\h20\m56\fs09 & $-$35\degr45\arcmin07\farcs3	& 4$\times$4 & Nominal & 11491 \\
1342251326 	& 2012-09-24	& NGC\,6334 I	  & 17\h20\m54\fs15 & $-$35\degr47\arcmin07\farcs4	& 4$\times$4 & Bright     & 9605   \\
1342214828 	& 2011-02-26	& NGC\,6334 `INT' & 17\h20\m48\fs10 & $-$35\degr45\arcmin42\farcs9	& Stare	     & Nominal  &  915 \\
1342214829	& 2011-02-26	& NGC\,6334 `OFF' & 17\h20\m39\fs04 & $-$35\degr43\arcmin43\farcs5	& Stare	     & Nominal & 915 \\
\hline
\end{tabular}
\tablefoot{
\tablefoottext{a}{Each individual observation has a hexagonal footprint spanning 3\arcmin\ in diameter. Their placement on the sky is illustrated in Fig.~\ref{fig:FTSfootprints}.} 
\tablefoottext{b}{An observation with a 4$\times$4 jiggle pattern on itself yields a spatially Nyquist sampled map, whereas a single `stare' observation probes positions spaced by roughly two full beam widths.}
} 
\end{table*}

\begin{figure}
	\resizebox{\hsize}{!}{\includegraphics{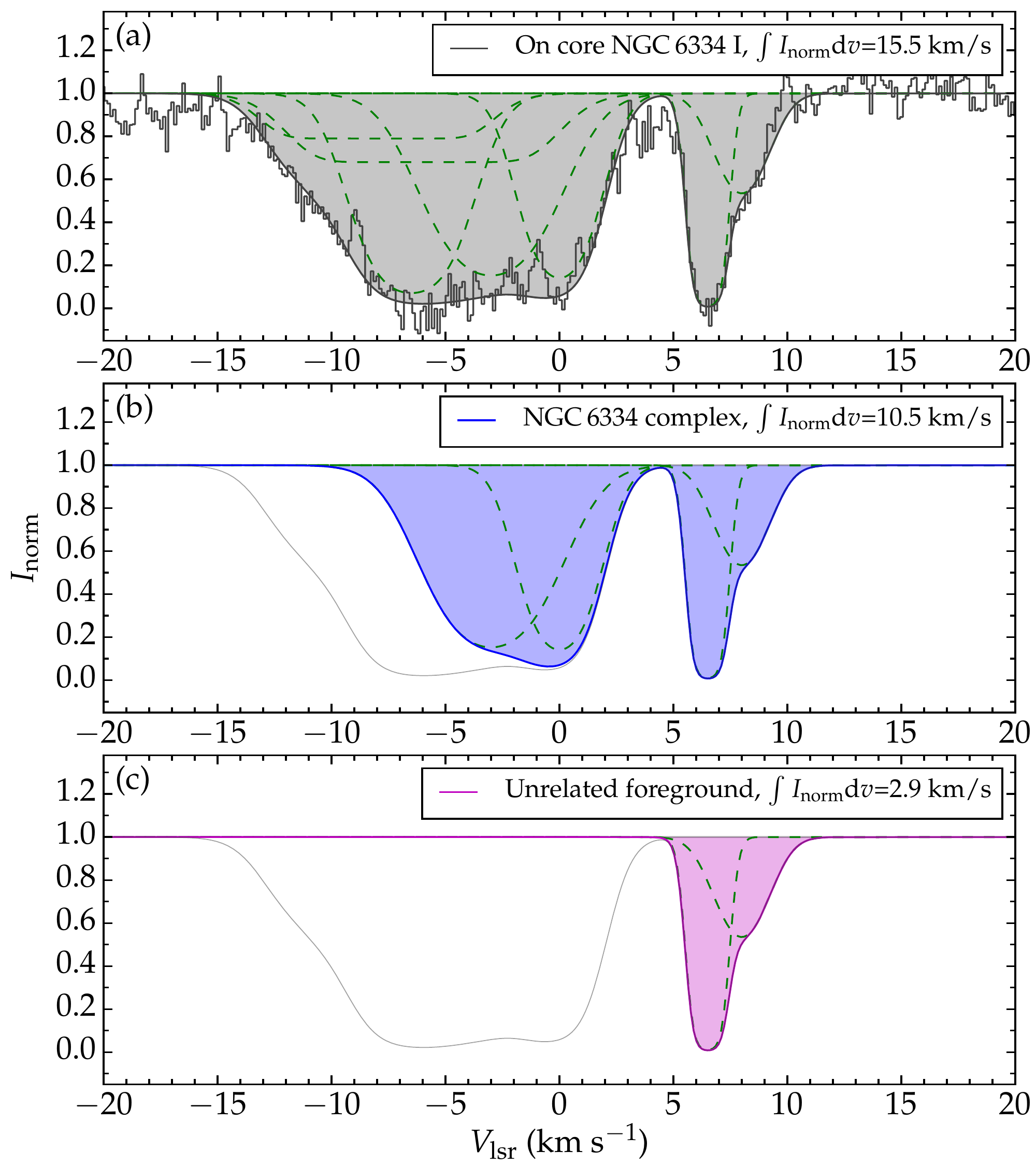}} 
	\caption{{\it Top panel}: velocity resolved spectral profile of HF absorption observed with HIFI toward core NG\,C6334~I (gray histogram). The seven individual model components are the same as in \citet{emprechtinger2012} and are drawn in dashed green lines, with their sum in solid gray. {\it Middle panel}: four foreground components expected to remain visible for lines of sight away from core I, but still on the NGC\,6334 molecular cloud complex (solid blue). {\it Bottom panel}: two foreground cloud components hypothesized to be unrelated to the NGC\,6334 complex (solid magenta).} 
	\label{fig:HFabscomp}
\end{figure}

The CH and HF signatures were observed toward NGC\,6334~I with the high spectral resolution spectrometer HIFI in single point mode \citep{degraauw2010}; its single pixel receiver did not provide any spatial information. In a \emph{map} of CH or HF absorption covering the region surrounding source I, one would expect to see a disentanglement of the different spatial extent of each velocity component as illustrated in Fig.~\ref{fig:HFabscomp}. Toward core I, the velocity resolved HF absorption signature, with a total equivalent width, $\int (1-I_\mathrm{norm}) \mathrm{d}V$, of 16~\kms, was modeled with seven components. At positions away from \mbox{core I}, but still on the NGC\,6334 filament, the equivalent width should diminish to 11~\kms\ representing the four foreground clouds, while at positions off of the cores and the filament, only the two foreground components that are more extended than the dense molecular cloud should be visible, and the equivalent width should drop to 3~\kms. 

This paper presents results from {\it Herschel} SPIRE iFTS spectral mapping observations toward a 6\arcmin$\times$3\farcm5 region surrounding cores I and I(N) in the NGC\,6334 star-forming complex. The observations are described in Sect.~\ref{sec:obs} and the resulting map of HF absorption depth is discussed in Sect.~\ref{sec:obsresults}. The signal is interpreted in Sect.~\ref{sec:analysis}, both in the context of foreground clouds and in that of freeze-out conditions in the dense cores. Conclusions are summarized in Sect.~\ref{sec:conclusions}.

\section{Observations and data reduction}
\label{sec:obs}

The spectral mapping observations used in this work were obtained as part of the `evolution of interstellar dust' guaranteed time program \citep{abergel2010} with the Spectral and Photometric Imaging Receiver \citep[SPIRE,][]{griffin2010} on board the {\it Herschel} space observatory \citep{pilbratt2010}. SPIRE's imaging Fourier Transform Spectrometer (iFTS) provides a jiggling observing mode that uses its 54 detectors to obtain Nyquist sampled spatial maps, covering the entire frequency range of the Spectrometer Long Wavelength (SLW, 447--1018~GHz) and the Spectrometer Short Wavelength (SSW, 944--1568~GHz) bands. The spectral resolution of 1.2~GHz corresponds to a resolving power $\nu/\Delta\nu\approx10^3$, roughly 300~\kms\ at the frequency of the HF 1--0 transition, 1232.5 GHz (243~\micron). 

Three partly overlapping, fully sampled SPIRE iFTS 4$\times$4 jiggle observations were performed in a total of nine hours of observing time, two centered on NGC\,6334 I and the third on NGC\,6334 I(N). A fourth, sparsely sampled observation, centered $\sim$2\arcmin\ northwest of core I, has considerable overlap with the combined area covered by the three other observations. This fourth observation is treated as an extra jiggle position in the gridding process described below. Finally, a fifth observation, also sparsely sampled, is centered 4\farcm5 northwest of core I and its footprint therefore has no overlap with our mapped area. At each jiggle position, four repeated scans of the FTS mechanism were executed in high spectral resolution mode. Details of the observations are summarized in Table~\ref{t:obs}. The placement of the different pointings described here is shown in Fig.~\ref{fig:FTSfootprints} in the Appendix. 

After inspection of the initial observations from February 2011, some detectors were found to suffer from saturation due to the bright emission toward source I. The observation toward source I was therefore repeated in `bright source mode' \citep{lu2014b} in September 2012 to obtain well calibrated spectra toward the brightest position. The majority of detector/jiggle combinations in the original observation point toward less bright regions and are therefore still useful in constructing the final map. 

The above SPIRE iFTS observations are processed with the `extended source' pipeline in HIPE 12.1.0 and the \texttt{spire\_cal\_12\_3} calibration tree, which includes the outer ring of partly vignetted detectors \citep{fulton2016}. The pipeline is interrupted at the pre-cube stage, before spectra from individual jiggle positions are gridded onto a rectangular spatial pattern. The spectrum for each jiggle position and each detector is visually inspected. Discarding all spectra that show excessive noise and/or irregular continuum shape (resulting from partial saturation) results in filtering 28 of the total of 919 SLW spectra (3\%) and 135 of the 1696 SSW spectra (8\%).  

The final processing step is the combination of the individual positions from each of the 19 (SLW) or 37 (SSW) detectors of each of the 55 jiggle positions into spectral cubes with square spatial pixels in Right Ascension and Declination coordinates. 
Due to the complex frequency dependence of the beam size and shape \citep{makiwa2013}, the native angular resolution of SPIRE iFTS observations varies non-monotonously across its frequency range. We use the convolution gridding scheme, which weighs each input value according to the distance from the target pixel center by means of a differential Gaussian kernel, with the aim of obtaining a cube with a constant effective reference beam of 43\arcsec\ FWHM. The pixel grid is identical for SLW and SSW, with square pixels measuring 17\farcs5$\times$17\farcs5. Gaussian convolution only results in a completely Gaussian target beam if the original beam is also well represented by a Gaussian shape. Such is the case for the entire SSW band and for low frequencies in SLW, but not for SLW frequencies between $\sim$700 and 1018~GHz \citep{makiwa2013}.

\begin{figure} 
	\resizebox{\hsize}{!}{\includegraphics{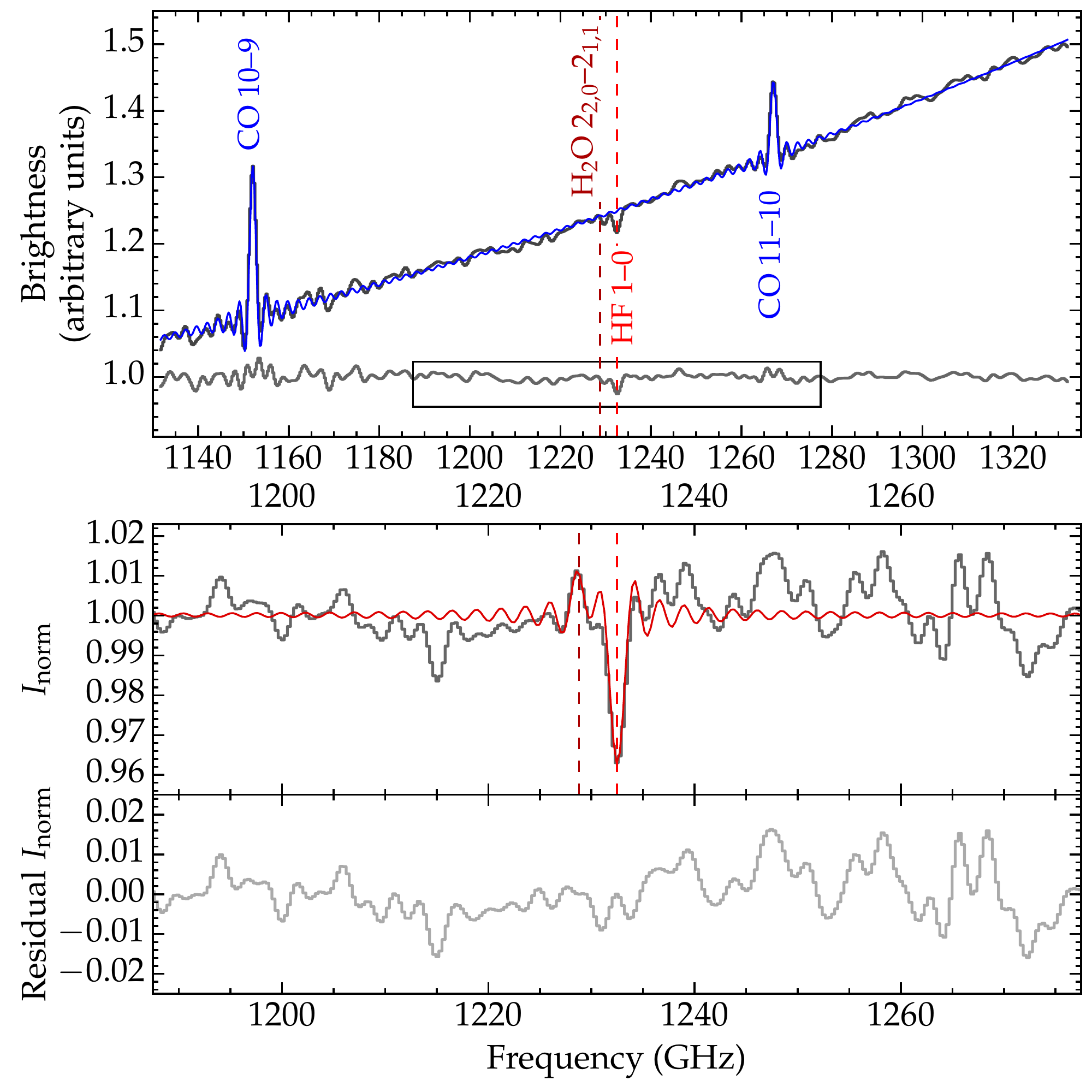}}
	\caption{Illustration of the procedure adopted to fit the spectrally unresolved, Sinc-shaped absorption line feature in the SPIRE iFTS cube. Rest frequencies of the relevant transitions are marked by vertical dashed lines. 
	{\it Top panel}: observed spectrum (black) toward one of the pixels in our map, fit to continuum and CO lines (blue), and the resulting ratio spectrum (gray). The rectangular box marks the axes limits of the next panel. 
	{\it Middle panel}: zoom of continuum-normalized spectrum (gray, same as top panel) and fit to the combined signature of the HF absorption line and the nearby \water\ emission line (red). 
	{\it Bottom panel}: residual remaining after subtracting the fit shown in the middle panel.}
	\label{fig:fittingprocess}
\end{figure}

\begin{figure}
	\resizebox{\hsize}{!}{\includegraphics{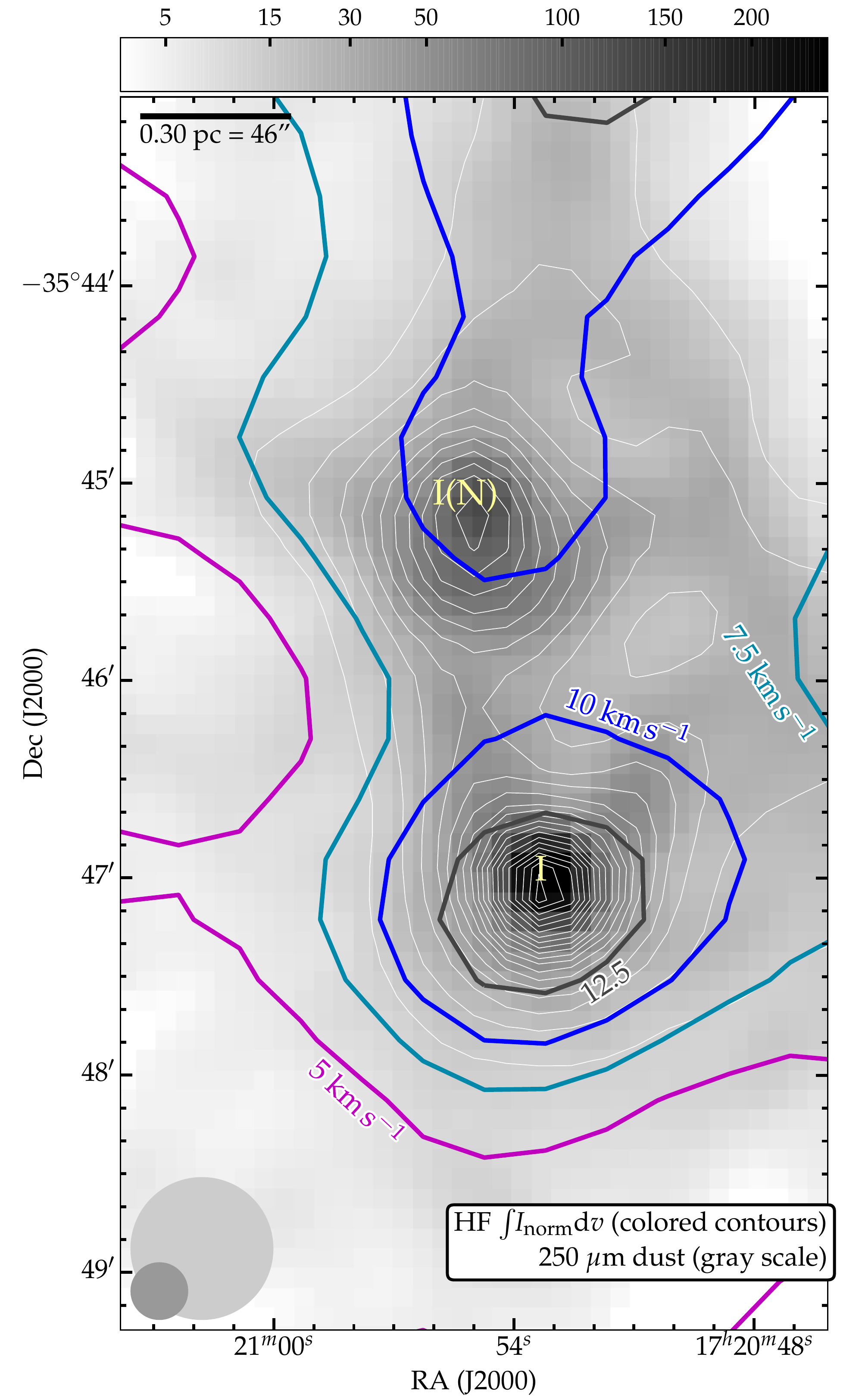}} 
	\caption{Map of equivalent width of the HF 1$\leftarrow$0 absorption signature measured with {\it Herschel} SPIRE iFTS (colored contours). The peak value in the center of the `12.5 \kms' contour is 16~\kms. The grayscale represents 250~\micron\ dust continuum emission measured with the SPIRE photometer ({\it Herschel} observation ID 1342239909, HIPE 13 standard pipeline, color bar stretched from 4 to 250 GJy\,sr$^{-1}$). The thin white contours are from the continuum as measured by the FTS, which closely follow the structure seen in the photometer grayscale map. Beam sizes of the convolved spectrometer cube and the photometer map are shown in the bottom left corner in light gray and dark gray circles, respectively. The scalebar in the top left corner indicates a projected length of 0.3~pc at a distance to the source of 1.35~kpc.   }
	\label{fig:HFabsmap}
\end{figure}

\section{Results}
\label{sec:obsresults}

The spectral cubes, as constructed in Sect.~\ref{sec:obs}, show a smooth dust continuum superposed with spectrally unresolved lines. The dominant line signal in the cube arises from the ladder of rotational transitions of CO ($J$=4--3 to 13--12). Early versions of the CO intensity maps of NGC~6334, based on subsets of the SPIRE iFTS observations used here, were presented in \citet{naylor2013} and \citet{makiwa2013osafts}. 
To retain the highest possible spectral resolution, we use unapodized FTS data, in which any unresolved spectral line has a Sinc-shaped profile. It is therefore important that one carefully fits and subtracts the Sinc-shaped profiles of nearby bright lines, to avoid any remaining sidelobes of strong lines affecting the apparent profile of the weak absorption line under study. 
This work focuses on absorption signatures of two hydride molecules, for which we use two spectral sections of the data cubes: 1132--1332~GHz from SSW and 760--935~GHz from SLW, chosen specifically to include the two CO emission lines closest to HF~1--0 at 1232.48~GHz \citep[rotational spectroscopy by][]{nolt1987} and \CHplus~1--0 at 835.08~GHz \citep[rotational transition frequency measured by][]{pearson2006}. We construct a script in HIPE \citep{ott2010}, derived from one of the post-pipeline analysis scripts provided by the SPIRE iFTS working group \citep{polehampton2015}, to fit a third order polynomial for the continuum simultaneously with the following lines: CO at 1152.0 and 1267.0~GHz (SSW), and CO at 806.7 and 921.8 and [\Catom] at 809.3~GHz (SLW). Knowing that the intrinsic width of CO lines in this region is only a few \kms\ \citep{zernickel2012}, lines are spectrally unresolved by SPIRE iFTS, and we adopt for each spectral line a Sinc profile with a fixed peak-to-first-zero-crossing width of 1.18~GHz. We then divide the observed spectrum by the fit of continuum and two/three Sinc lines to obtain a continuum-normalized spectrum, $I_\mathrm{norm}$. This process, illustrated in the top panel of Fig.~\ref{fig:fittingprocess}, is repeated for each spatial pixel in the cube. 

In the resulting continuum-normalized cube around 1232~GHz, we fit two Sinc functions to the absorption profile of HF 0$\leftarrow$1 at 1232.48~GHz and the nearby emission line of \water~2$_{2,0}$$\rightarrow$2$_{1,1}$ at 1228.79~GHz (Fig.~\ref{fig:fittingprocess}, middle panel). In the continuum-normalized cube around 835~GHz, we fit three Sinc functions to the absorption profile of \CHplus\ 0$\leftarrow$1 at 835.08~GHz and \thCO\ emission lines at 771.18 and 881.27~GHz. 
The maps of equivalent width of the HF and \CHplus\ absorption depth, in units of Hz, are multiplied by the ratio of the speed of light, c, and the observed frequency, $\nu_\mathrm{obs}$, to convert to \kms\ units: c/$\nu_\mathrm{obs}$=\pow{2.43}{-7}~\kms\,Hz$^{-1}$ for HF, and \pow{3.59}{-7}~\kms\,Hz$^{-1}$ for \CHplus. Line fits are rejected if the signal-to-noise ratio is lower than 2 and/or the fitted line center is more than half a resolution element from the line's expected frequency at $-8.3$ \kms\ \vlsr\ \citep{vanderwiel2010}. The resulting map of HF equivalent width in Fig.~\ref{fig:HFabsmap} reveals the spatial distribution of the HF absorption feature, detected in 81\% (signal-to-noise$>$2) or 72\% (signal-to-noise$>$3) of all the pixels in the 6\arcmin$\times$3\farcm5 map coverage. See also the signal-to-noise map in Fig.~\ref{fig:SNmapHF}. 

Uncertainties are calculated from the spectral rms noise in the continuum-normalized cubes in two 20~GHz ranges surrounding the HF absorption line. For \CHplus, the frequency ranges for calculating noise are composed of a 10~GHz section below and a 30~GHz section above the \CHplus\ frequency, to avoid incorporating residual from the fit to the blended CO~7--6 and [\Catom]~$^3$P$_2$--$^3$P$_1$ lines at 806 and 809 GHz. The \mbox{1-$\sigma$} noise on the equivalent width is obtained by multiplying the unitless spectral rms with the instrumental line width of 1.18~GHz $\times$ c/$\nu_\mathrm{obs}$. Noise values are variable across the maps, in the range 1.2--3.4 \kms\ for HF and 1.4--7 \kms\ for \CHplus. We do not include the following contributions to the uncertainty. Firstly, any multiplicative effects such as those of the absolute intensity calibration \citep{benielli2014,swinyard2014} are divided out by normalizing the spectra to the local continuum. Secondly, additive uncertainties in the continuum level offsets are $\sim$\pow{0.5}{-19} \Wsqm\ for SLW and $\sim$\pow{2}{-19} \Wsqm\ for SSW \citep{swinyard2014,hopwood2015}. These values are negligible compared to the brightness of the continuum in our cube, which exceeds the offset uncertainty by a factor of a few hundred even in the faintest outer regions. 

Spectra from the `OFF' observation, pointed just northwest of the mapped area shown in Fig.~\ref{fig:HFabsmap}, are also inspected at the HF frequency, but no convincing detections are found. The spectra from individual detectors in the OFF observation exhibit rms noise values between 4 and 10 \kms, with a median of 6 \kms. This noise is considerably higher than that in the cube pixels based on the other four observations combined, in which each pixel encompasses at least four, but typically more than eight individual detector pointings. The lack of HF absorption detections in the OFF position is thus consistent at the 2-$\sigma$ level with an HF absorption depth $\lesssim$12~\kms\ in the area 3--6\arcmin\ northwest of source I, i.e., absorption depths could be anywhere in the range shown in our mapped area, except the central 40\arcsec\ around source I itself where the strongest absorption is seen. 

To rule out contamination of the HF signature by other spectral lines within SPIRE's spectral resolution element, we inspect the high spectral resolution spectrum toward the position of the chemically rich core NGC\,6334~I \citep{zernickel2012}, observed as part of the spectral survey key program CHESS \citep{ceccarelli2010} using the HIFI spectrometer \citep{degraauw2010,roelfsema2012}. The only spectral lines detected by HIFI in a frequency span of $\pm$2~GHz around the HF frequency are four marginally detected methanol emission lines (A.~Zernickel, private communication, Jun.~2014) together amounting to $<$0.4~\kms\ in equivalent width. The possible methanol contaminations for the measured HF absorption depth are therefore contained within the uncertainty for our HF equivalent widths quoted above. The effect of the \water\ emission line at 1229~GHz (see also above) could be more significant: at the brightest position, toward core I, the water line is as bright as one third of the deepest HF absorption. As described above and shown in Fig.~\ref{fig:fittingprocess}, the effects of the water line on the HF absorption profile are taken into account by applying a simultaneous fit of these two lines, separated in frequency by three times the SPIRE instrumental line width. 

We also detect the signature of \CHplus\ 1$\leftarrow$0 absorption at 835.08~GHz in the spectral map from the SLW array. We refrain from interpreting its signal here for the following reasons. 
Firstly, the signal-to-noise ratio in the \CHplus\ absorption map is much lower than that in the HF map (see Fig.~\ref{fig:SNmapHF} and \ref{fig:SNmapCHplus}), resulting in signal-to-noise$>$2 detections in only 46\% of the mapped pixels. For completeness, the distribution of detected \CHplus\ absorption is shown in the Appendix in Fig.~\ref{fig:SNmapCHplus}. Importantly, around the position of source I, there is no confident detection to be compared with heterodyne observations from \citet{zernickel2012}. Only a few isolated pixels near that position have detections of \CHplus, but at a signal-to-noise of $<$3. 
Secondly, the spectrally resolved HIFI spectrum toward source~I (A.~Zernickel, private communication, Jun.~2014) show seven distinct emission line features due to methanol at frequencies within 0.6~GHz of the \CHplus\ line, i.e., half of the SPIRE spectral resolution. The combined intensity of these lines is sufficient to compensate more than half of the \CHplus\ absorption observed by HIFI, and they therefore severely contaminate the spectrally unresolved profile of \CHplus\ absorption in the SPIRE spectrum. In fact, these methanol emission lines could be the cause of the weak detection at the position of core I with SPIRE's modest spectral resolution. Thirdly, the \CHplus\ transition falls in a frequency range in which the beam profile of the SPIRE iFTS is non-Gaussian in shape \citep{makiwa2013}, complicating the map convolution and the interpretation of any spatial structure. 

Our SPIRE spectroscopic data also show evidence for detections of NH$_2$ at 952.6 and 959.5 GHz, 
and NH at 974.5~GHz 
and at 1000 GHz. 
However, compared to HF, it is more challenging for astrochemical models to explain the observed abundances of N-bearing hydrides \citep[e.g.,][]{persson_cm2012}, and the analysis of line features is complicated by hyperfine structure within rotational transitions \citep[cf.~spectrally resolved detections with HIFI by][]{zernickel2012}. For these two reasons, we refrain from interpreting the NH and NH$_2$ absorption depth maps in this paper, but for completeness they are shown in Figs.~\ref{fig:SNmapNH} and \ref{fig:SNmapNH2}. 
The OH$^+$ doublet at 909.05 and 909.16~GHz is also within the frequency range covered by the SPIRE iFTS, but the bright methanol emission line at 909.07~GHz apparent in the HIFI spectrum published by \citet{zernickel2012} would make analysis of the blended OH$^+$ signature in the SPIRE spectrum impossible.

\section{Analysis and discussion} 
\label{sec:analysis}

\subsection{HF optical depth, equivalent width, and column density}
\label{sec:eqwidth}

An absorption line is saturated when its depth reaches zero, i.e., absorbing all continuum photons in any specific spectral channel. HIFI observations seem to show that the combined absorption feature of HF due to the NGC\,6334~I envelope, hot cores, and the foreground cloud at $-3.0$~\kms\ is saturated between \vlsr\ of $-7$ and $-4$~\kms\ (Fig.~\ref{fig:HFabscomp}a). The individual components, however, do not reach 100\% absorption. Of all seven components, the cloud at $+6.5$~\kms\ comes closest to having saturated absorption in HF. As already noted by \citet{emprechtinger2012}, even this component is only marginally optically thick, evidence for which is provided by the line width of the same component in the optically thin tracer CH \citep{vanderwiel2010} which is the same (1.5~\kms) as for HF. All other components are believed to be optically thin \citep{emprechtinger2012}. 

Absorption line depth is converted into optical depth using $I_\mathrm{norm}=\mathrm{e}^{-\tau_\nu}$. With the caution of one of the seven components being marginally saturated, we take the optically thin limit to relate optical depth integrated over the line profile, \taudv, to column density, $N_\mathrm{HF}$, following \citet{neufeld2010b}: 
\begin{equation}
\label{eq:tautocolumn}
\int \tau_{\nu,\mathrm{HF}} \mathrm{d}V = \frac{A_\mathrm{ul} g_\mathrm{u} \lambda^3}{8\pi g_\mathrm{l}} N_\mathrm{HF} = \powm{4.16}{-13} \, [\mathrm{cm}^2 \, \mathrm{km}\, \mathrm{s}^{-1}]\ N_\mathrm{HF} , 
\end{equation} 
where $A_\mathrm{ul}$ is the Einstein $A$ coefficient for spontaneous emission, \pow{2.42}{-2} s$^{-1}$ for HF 1--0 \citep{pickett1998}, $g_\mathrm{u}$=3 and $g_\mathrm{l}$=1 are the statistical weights of rotational levels $J$=1 and $J$=0, respectively, and $\lambda$ is the wavelength of the transition, 243.24 \micron. In addition to the optically thin limit, Eq.~(\ref{eq:tautocolumn}) assumes that all HF molecules are in the rotational ground state, a fair assumption given its high $A_\mathrm{ul}$. 

The conversion from $I_\mathrm{norm}$ to $\tau_\nu$ follows a linear relation for low values of $\tau_\nu$: 
\begin{eqnarray}
\nonumber
\tau_\nu & = & -\ln(I_\mathrm{norm}) \\ 
 & \approx & 1-I_\mathrm{norm} \qquad  [\mathrm{for}\ I_\mathrm{norm} \approx 1]. 
\end{eqnarray}
This relation holds to within $\sim$10\% for $\tau_\nu<0.2$ (line absorbs up to 20\% of the continuum), but $\tau_\nu/(1-I_\mathrm{norm})$ is already 1.2 at 40\% absorption, rising to 2 at 80\% absorption. 
The integrated optical depth is therefore systematically underestimated for a spectrally unresolved line that is smeared out over a velocity range wider than its intrinsic profile.\footnote{The logarithm operator gives a disproportionally higher weight to more strongly absorbed channels. For example, a 20 \kms\ wide feature with constant 80\% absorption ($I_\mathrm{norm}$=0.20) and a smeared-out version of 320 \kms\ wide at 5\% absorption ($I_\mathrm{norm}$=0.95) both have the same equivalent width, \mbox{$\int(1-I_\mathrm{norm}) \mathrm{d}V$} = 16~\kms. The latter, however, yields a $\int \tau_\nu \mathrm{d}V$ that is smaller by a factor 2. The effect is larger yet for line profiles that come even closer to full absorption.}  
Since the HF line is spectrally unresolved in our SPIRE spectra, a column density derived from these observations would merely constitute a lower limit to the true column density. 
Contrary to the optical depth, the absorption depth integrated over the line profile, i.e., the equivalent width of the absorbed `area' below $I_\mathrm{norm}$=1, is conserved regardless of the spectral resolution. This is confirmed by the matching equivalent width values measured by HIFI and SPIRE toward core I: \mbox{$\int(1-I_\mathrm{norm}) \mathrm{d}V$} is 15.7 and 16.4~\kms, respectively, with uncertainty margins of $\sim$1~\kms\ in both cases. 
In the remainder of this paper, we analyze HF absorption depth measured with SPIRE based exclusively on the conserved quantity, equivalent width, \mbox{$\int(1-I_\mathrm{norm}) \mathrm{d}V$}. When deriving optical depths and column densities, we rely exclusively on spectrally resolved profiles such as that in \citet{emprechtinger2012}.

\subsection{Distribution of HF absorbing clouds toward NGC\,6334} 
\label{sec:absclouds}

The range of HF equivalent width values in the map shown in Fig.~\ref{fig:HFabsmap} can be divided into three regimes: (a) $>$12~\kms, only occurring toward the position of core I; (b) 8--12~\kms, spatially consistent with the larger scale filament in which cores I and I(N) are embedded; and (c) $\lesssim$5~\kms, exclusively localized at projected distances $>$0.6~pc from the cores and the connecting filament. The non-detection of HF in the `OFF' observation (see Sect.~\ref{sec:obsresults}), sparsely sampling the area just northwest of our map, is consistent with regimes (c) or (b). 

\subsubsection{Distinguishing foreground from dense star-forming gas}
\label{sec:distinguishforeground}

We interpret the three regimes in the context of the velocity resolved HF spectrum published by \citet{emprechtinger2012}, who identified seven individual physical components responsible for the HF absorption toward NGC\,6334~I: the dense envelope at \vlsr=$-6.5$~\kms, two compact subcores at $-6.0$ and $-8.0$~\kms, and four foreground layers at $-3.0$, $0.0$, $+6.5$, and $+8.0$~\kms. The spectral signature of each of these components is reproduced in our Fig.~\ref{fig:HFabscomp}a. Regime (a) requires all seven components to explain the total equivalent width of HF. The two other panels in Fig.~\ref{fig:HFabscomp} represent adaptations of the model from \citet{emprechtinger2012} with progressively fewer absorption components taken into account. In Fig.~\ref{fig:HFabscomp}b, regime (b) is explained by the superposition of four absorbing foreground clouds, discarding the components associated with the envelope and subcores of NGC\,6334~I. We highlight that the HF absorption depth observed toward the dense star-forming envelope I(N) is consistent with regime (b). Differences in HF content between the dense envelopes I and I(N) are discussed in more detail in Sect.~\ref{sec:freezeout}. 
Finally, in Fig.~\ref{fig:HFabscomp}c, we show that regime (c) is consistent with a model composed of just two specific foreground clouds, namely those at $+6.5$ and $+8.0$~\kms\ \citep{vanderwiel2010,emprechtinger2012}. 

In contrast with the detailed study of the HF profile in \citet{emprechtinger2012} and in this work, the spectral survey paper by \citet{zernickel2012}, analyzing $\sim$4300 individual spectral line features, uses a simplified model which explains the HF absorption with only three components. The two approaches are not inconsistent, but the latter paper groups components together as: (1) the NGC\,6334~I envelope and two subcores, (2) two foreground clouds that are kinematically close to the NGC\,6334 complex, and (3) two other foreground clouds with larger \vlsr\ offsets \citep[cf.~Table~1 in][]{emprechtinger2012}. In this three-component model, regime (a) would be explained by groups (1)+(2)+(3), regime (b) by groups (2)+(3), and regime (c) by only group (3). 

\begin{figure}
	\resizebox{\hsize}{!}{\includegraphics{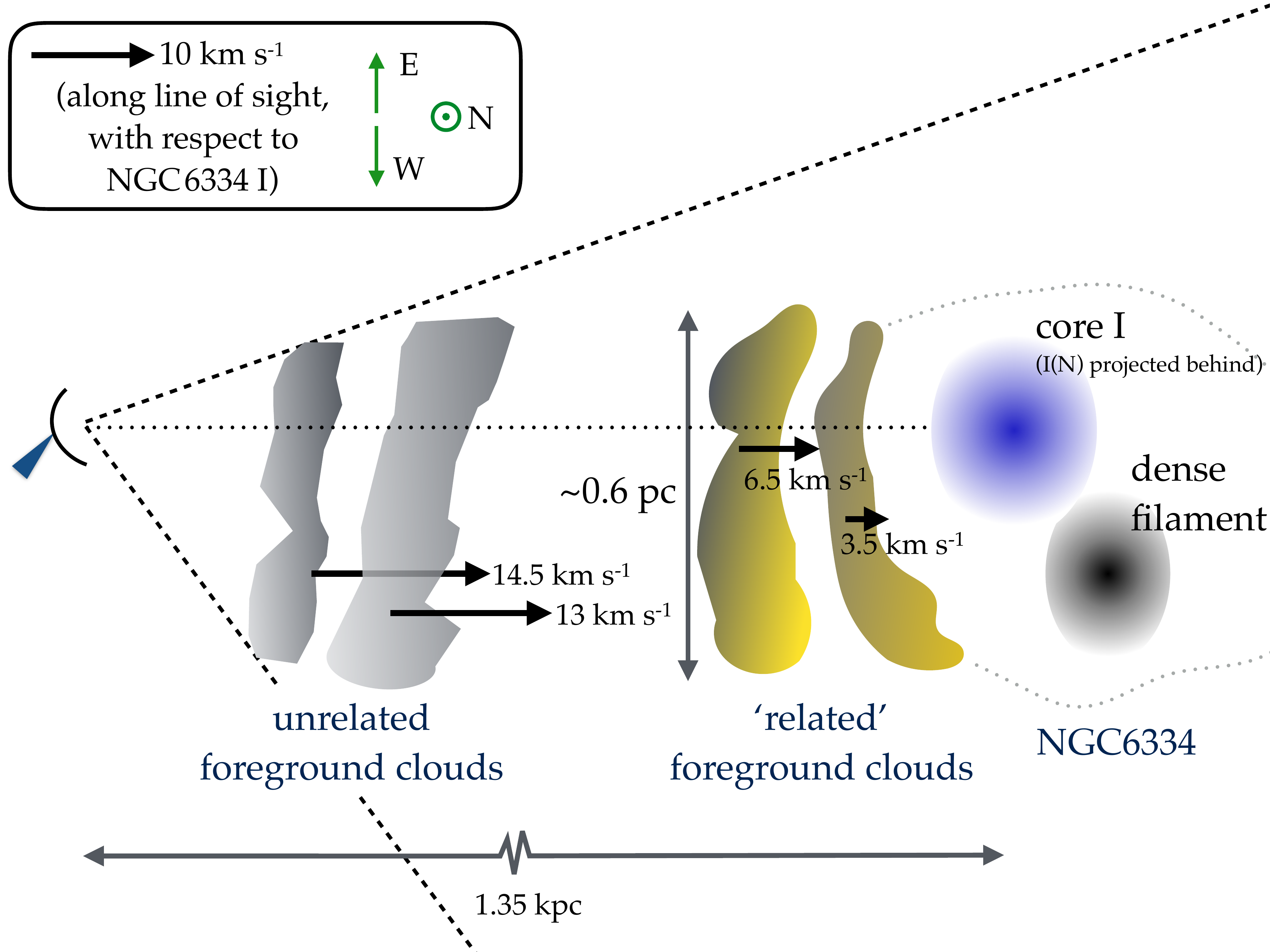}} 
	\caption{Sketch of the location and extent of the foreground clouds. Note the broken scale in the horizontal direction. Velocity magnitudes are indicated with black arrows, shifted such that the core NGC\,6334~I, at \vlsr=$-6.5$~\kms, is at rest in this frame. From left to right, they represent the components referred to in the text as the $+8.0$, $+6.5$, $+0.0$, and $-3.0$~\kms\ clouds. The dotted line represents the line of sight from the observer to NGC\,6334~I. The dashed lines indicate the angular extent of clouds on the far foreground. The color differential between core I and the filament indicates the elevated temperature in core I, preventing freeze-out of HF molecules.}
	\label{fig:geometrysketch}
\end{figure}

The combination of our HF absorption depth map with the previous, single-pointing, velocity resolved HF spectrum now reveals a three-dimensional picture of the layers of absorbing gas toward the NGC\,6334 complex. Our interpretation of the relative positions of the foreground layers is sketched in Fig.~\ref{fig:geometrysketch}. With the exception of the direction toward core I, the HF absorption depth at all other positions in the map can be explained by (a subset of) the four extended foreground clouds. 

\subsubsection{Relation of foreground clouds to NGC\,6334} 
\label{sec:foregroundrelation}

\citet{vanderwiel2010} used a single-pointing {\it Herschel} HIFI observation of CH rotational line absorption, coupled with results from OH hyperfine line absorption measurements at radio wavelengths \citep{brooks2001}, to suggest that the clouds at \vlsr=$-3.0$ and $+0.0$~\kms\ are associated with the NGC\,6334 complex, while the remaining two velocity components, at $+6.5$ and $+8.0$~\kms, originate in foreground clouds farther away from the NGC\,6334 complex. This interpretation is consistent with the first two components being seen exclusively in regime (b) of the HF equivalent width map and the latter two components being spread over a more extended area of regimes (b) and (c) combined. Therefore, the spatial distribution of HF absorption measured with SPIRE iFTS supports the previous hypothesis that the $-3.0$ and $+0.0$~\kms\ clouds are associated with NGC\,6334, since they have a spatial morphology that closely follows the dense molecular cloud traced by the 250~\micron\ dust emission (Fig.~\ref{fig:HFabsmap}), i.e., a region roughly \mbox{2--3\arcmin} in east-west extent and stretching along the north-south direction. The vertical extent of the `related' foreground clouds depicted in Fig.~\ref{fig:geometrysketch} is drawn directly from the observed east-west extent of these component in the plane of the sky.  

Besides the diatomic hydride species HF, CH, and OH, a subset of our absorbing clouds also exhibit absorption lines due to \water\ \citep{emprechtinger2010,vandertak2013a}, \waterp\ \citep{ossenkopf2010b}, and H$_2$Cl$^+$ \citep{lis2010a}. Absorption components in OH$^+$ and \waterp\ detected toward both cores by \citet{indriolo2015a} peak at \vlsr=$-2$ and $+3$~\kms. The former could be a blend of the $-3.0$ and $+0.0$~\kms\ clouds seen in CH and HF, but the latter is inconsistent with any of our components. 
These lines were all detected in observations with the single-pixel, high spectral resolution {\it Herschel} HIFI spectrometer, in some cases toward both individual protostellar cores. It is interesting that ionized species, \waterp\ and H$_2$Cl$^+$, are only detected at \vlsr\ values that match the $-3$ and $+0$~\kms\ clouds. Chemical models tailored to halogen hydrides \citep{neufeld2009b} indicate that cation species become abundant under the influence of strong UV radiation. We hypothesize that the presence of \waterp\ and H$_2$Cl$^+$ is further evidence for the physical proximity of these two clouds to the massive protostars and the \Hii\ region embedded in one of the dense cores. 

Another ionized species that has been detected in spectrally resolved observations toward NGC\,6334 is \CHplus. 
While a velocity resolved observation of HF exists only toward the position of core I, \CHplus~1--0 has been observed with HIFI toward both cores I and I(N), the latter as part of the WISH program \citep{vandishoeck2011,benz2016}. Toward source I, the \CHplus\ profile looks very similar to the HF profile, amounting to a total equivalent width of $\sim$20~\kms. In comparison, the \CHplus\ observation toward core I(N)\footnote{Estimate based on the HIFI spectrum downloaded from the \mbox{\it Herschel} Science Archive, observation ID 1342214306, processed with HIPE pipeline version 13.} reveals an equivalent width of only $\sim$13~\kms. The majority (85\%) of the reduced absorption toward core I(N) relative to core I falls in the $[-15,5]$\,\kms\ range, as expected if the missing components are the envelope and subcore components at \vlsr=$-6.5$, $-6.0$, and $-8.0$~\kms\ (cf.~Fig.~\ref{fig:HFabscomp}), i.e., those components that occur only in regime (a) in our HF absorption map. As mentioned above in Sect.~\ref{sec:obsresults}, the SPIRE iFTS spectral cube has too low signal-to-noise near 835~GHz to make a meaningful comparison with the \CHplus\ signature detected by this instrument. 

The foreground clouds at \vlsr=$+6.5$ and $+8.0$~\kms, supposedly unrelated to the NGC\,6334 dense filament \citep{brooks2001,vanderwiel2010}, have a combined HF equivalent width that is entirely consistent with the observation in this work that regime (c) is spatially extended beyond the dense filament. An unrelated set of foreground cloud(s) (the two leftmost clouds in Fig.~\ref{fig:geometrysketch}) are likely to have a larger angular extent than the background continuum source (dashed lines in Fig.~\ref{fig:geometrysketch}), despite a possibly modest linear size. This explains why a minimum level of HF equivalent width of $\sim$3~\kms\ is detected not just toward the NGC\,6334 filament, but throughout the extent of our 6\arcmin$\times$3\farcm5 map (Figs.~\ref{fig:HFabsmap} and \ref{fig:SNmapHF}). 

All four foreground clouds detected in HF and CH are redshifted (\vlsr$\geq$$-3$~\kms) with respect to the \vlsr\ of the part of NGC\,6334 cloud near cores I and I(N) (around $-5$~\kms, based on \HCOplus\ observations by \citealt{zernickel2013}). Thus, the absorbing gas clouds are moving \emph{toward} NGC\,6334, instead of following the Galactic rotation, which at $\ell=351\degr$ yields exclusively negative line-of-sight velocities for sources between Sun and NGC\,6334, i.e., approaching the standard of rest of stars in the Solar neighborhood. 
We therefore conclude that, in addition to the gas flows within the dense gas \emph{along} the filament's long axis \citep{zernickel2013}, gas may also be accreting onto the filament in the perpendicular direction. 
We also note that, whereas \citet[][their Sect.~3.7 and Table 5]{indriolo2015a} put the \OHp\ and \waterp\ absorption clouds toward both cores I and I(N) at the distance of 1.35~kpc of the NGC\,6334 cloud, their location along the sight line toward NGC\,6334 is in fact poorly constrained. Recognizing that there must be peculiar motions at play, deviating from the `rigid' Galactic rotation curve, these clouds could in fact be anywhere between the local arm of the Milky Way and the Sagittarius arm that harbors the NGC\,6334 complex. 

For the two foreground clouds related to NGC\,6334, at $-3$ and $+0$~\kms, we calculate their total mass by multiplying the sum of their column densities with a rough estimate of the area covered by this component of 160$\times$270\arcsec\ (regime (b) in Fig.~\ref{fig:HFabsmap}), corresponding to 1.5 pc$^2$ at the distance of NGC\,6334. Depending on the choice of fiducial \HH\ tracer, either taking $N_\mathrm{CH}$ from \citet{vanderwiel2010} and $N_\mathrm{CH}$/$N_\mathrm{H_2}$=\pow{(2.1--5.6)}{-8} from \citet{sheffer2008}, or $N$(HF) from \citet{emprechtinger2012} and $N_\mathrm{HF}$/$N_\mathrm{H_2}$=\pow{(0.5--1.4)}{-8} from \citet{indriolo2013a}, this calculation yields a mass in the range 37--98 \Msun\ or  69--191 \Msun, respectively. From symmetry arguments, a similar reservoir of additional gas is expected to lie behind the NGC\,6334 dense cloud. This means that a significant total mass of several hundred \Msun\ could be on its way to accreting onto the dense cloud NGC\,6334 near ($<$0.3~pc) the embedded cores I and I(N). This gas reservoir has escaped detection so far, because it does not appear in traditional gas tracers such as CO, \HCOplus, and CS.
While there is evidence that the $-3.0$ and $+0.0$~\kms\ foreground clouds are closer to NGC\,6334 than the $+6.5$ and $+8.0$~\kms\ clouds, there is no direct metric of the geometrical distance along the line of sight from each cloud to the dense filament and cores. Therefore, we refrain from speculating about accretion time scales of even the `related' clouds, since this would rely on unsupported assumptions on relative distances. 

\subsubsection{Spatial distribution of HF toward other Galactic sight lines}
The only other published work of a spatial map of HF absorption so far is that toward \object{Sgr~B2(M)} by \citet{etxaluze2013}, using data also obtained with {\it Herschel} SPIRE iFTS. A direct comparison of measured absorption line depths is complicated by the choice of \citet{etxaluze2013} to present signal in terms of integrated optical depth, apparently without taking into account the systematic underestimation of optical depths derived from spectrally unresolved measurements, as discussed in Sect.~\ref{sec:eqwidth}. Nonetheless, it appears that the total HF absorption toward Sgr~B2(M) is about an order of magnitude stronger than toward NGC\,6334~I. \citet{etxaluze2013} find a variation of HF absorption depth of only a factor $\sim$2 across the $\sim$2\farcm5 mapped area, significantly less variation than in our fully sampled map toward NGC\,6334~I. Any intrinsic variation may have been partially smoothed by the interpolation process that was applied in \citet{etxaluze2013} to construct a map from the spatially undersampled SPIRE iFTS observation. More importantly, the line of sight toward Sgr~B2, close to the Galactic Center, crosses many more spiral arms than that toward NGC\,6334. Evidence of this is found for example by \citet{qin2010}, who detect a total of 31 individual velocity components in CH rotational ground state absorption toward Sgr~B2(M). The 30 foreground clouds, associated with the various intervening Galactic arms, amount to a total CH column density of \pow{1.7}{15}~\psqcm, with Sgr~B2(M) itself adding a component of only \pow{0.5}{15}~\psqcm\ \citep{qin2010}. This is consistent with many sheets of foreground gas together creating a roughly uniform cover of absorbing gas spanning at least a few arcminutes on the sky. Comparatively little additional absorption is contributed by the massive molecular cloud Sgr~B2 and the embedded cores in the background, which explains the lack of variation in HF absorption depth seen toward Sgr~B2 by \citet{etxaluze2013}. In contrast, our map of HF absorption toward NGC\,6334 reveals a mix of components due to foreground clouds and the star-forming envelope and cores within NGC\,6334, which we are able to disentangle owing to the complementary, velocity resolved spectra of CH and HF obtained with {\it Herschel} HIFI \citep{vanderwiel2010,emprechtinger2012}. 

In addition, we highlight the discovery by \citet{lopez-sepulcre2013a} of a foreground cloud toward the intermediate-mass star-forming core \object{\mbox{OMC-2 FIR\,4}}, based on single-pointing HIFI spectra of HF and other hydride molecules. Several oxygen-bearing hydrides show absorption exclusively at a blueshifted velocity relative to the background protostellar core. The same foreground cloud was later also identified in H$_2$Cl$^+$ by \citet{kama2015a}. The HF profile shows absorption at the same blueshifted velocity, but shows additional evidence for a second absorption component that matches the \vlsr\ of the protostellar core \citep{lopez-sepulcre2013a}. 
Instead of kinematical and morphological arguments such as those used in this work to determine the physical location of foreground clouds toward NGC\,6334, \citet{lopez-sepulcre2013a} use detailed photochemical modeling to infer proximity of their OMC-2 foreground gas to a source of copious far-UV radiation. With that radiation source assumed to be the trapezium cluster of OB stars, it is concluded that the absorbing slab is physically connected to OMC-1. In an attempt to study the spatial distribution of this absorbing OMC-1 `fossil' slab, we have searched archival SPIRE iFTS data toward \mbox{OMC-2 FIR\,4} ({\it Herschel} observation ID 1342214847) for signatures of HF at 1232.5~GHz (and \CHplus\ at 835.1~GHz), but find no detections in any of the 37 (and 19) SSW (and SLW) detectors in the $\sim$3\arcmin\ footprint.

\subsection{Freeze-out of HF in envelopes of dense cores}
\label{sec:freezeout}

The HF equivalent width of 15.9$\pm$1.4~\kms\ measured at the position of NGC\,6334~I in our SPIRE map is explained in Sect.~\ref{sec:absclouds} by invoking the superposition of four absorption components in the foreground and three associated to the dense core I itself \citep[see Fig.~\ref{fig:HFabscomp}, and][]{emprechtinger2012}. A striking feature of our HF absorption map is the lack of additional absorption toward the position of core I(N), where the HF equivalent width of only 10.9$\pm$1.1~\kms\ can be explained by the four foreground clouds alone, adding up to 10.5~\kms\ (Fig.~\ref{fig:HFabscomp}b). In contrast, adding even just an envelope component similar to that of core I sums up to 13.7~\kms\ (not counting the two subcore components), which is inconsistent with the observation toward core I(N). Since the envelope of core I(N) is more massive than that of core I, but has a similar size \citep[see model fits in][]{vandertak2013a}, the total gas column density toward core I(N) should be higher. Therefore, the lack of HF absorption associated to the I(N) core is not due to the difference in total (\HH) column. Instead, we hypothesize that HF is primarily frozen out onto dust grains in core I(N), while HF is in the gas phase in core I. 

To support the hypothesis of HF being depleted from the gas phase in core I(N), we set up a rudimentary model based on the following ingredients. We take the spherically symmetric physical structure, i.e., radial profiles of density and temperature, of the envelopes of NGC\,6334~I and I(N) as fitted to submillimeter dust continuum maps and the far-infrared / submillimeter spectral energy distribution  \citep{vandertak2013a}. We then calculate, at every radial point, the timescales for adsorption (freeze-out) and desorption (evaporation) of HF molecules onto dust grains. Following, e.g., \citet{rodgers2003} and \citet{jorgensen2005a}, we assume that thermal desorption is the dominant mechanism that drives molecules from the grain surface back into the gas phase, and are left with the balance between adsorption rate: 
\begin{equation}
\lambda(n_\mathrm{H}, T_\mathrm{gas}) = \powm{4.55}{-18} \left( \frac{T_\mathrm{gas}}{m_\mathrm{HF}} \right)^{0.5} n_\mathrm{H} \qquad [\mathrm{s}^{-1}] , 
\label{eq:freezerate}
\end{equation}
and desorption rate:
\begin{equation}
\xi(T_\mathrm{dust}) = \nu_\mathrm{vib} \exp\left( -\frac{E_\mathrm{b,HF}}{k\,T_\mathrm{dust}} \right)   \qquad [\mathrm{s}^{-1}] .
\label{eq:desorbrate} 
\end{equation} 
Here, \Tgas\ and \Tdust\ are the temperatures of gas and dust, assumed to be equal as in the modeling of \citet{vandertak2013a}, $m_\mathrm{HF}$ is the molecular weight of HF (20), $n_\mathrm{H}$ is the density of hydrogen nuclei, $\nu_\mathrm{vib}$ is the vibrational frequency of the HF molecule in its binding site, for which we adopt $10^{13}$~s$^{-1}$, $k$ is the Boltzmann constant, and $E_\mathrm{b,HF}$ is the binding energy of HF to the dust grain surface. It has previously been inferred by \citet{philips2010} that a density of $\sim$$10^5$~\pccm\ allows HF to condense onto dust grains, whereas densities of $\sim$$10^3$~\pccm\ more typical for diffuse gas are too low for HF freeze-out to occur. In our case, the density $n_\mathrm{H}$ -- which incidentally exceeds $10^5$~\pccm\ at almost all radii in the envelopes of I and I(N) -- enters directly into Equation~\ref{eq:freezerate} to govern the adsorption rate.

\begin{figure} 
	\resizebox{\hsize}{!}{\includegraphics{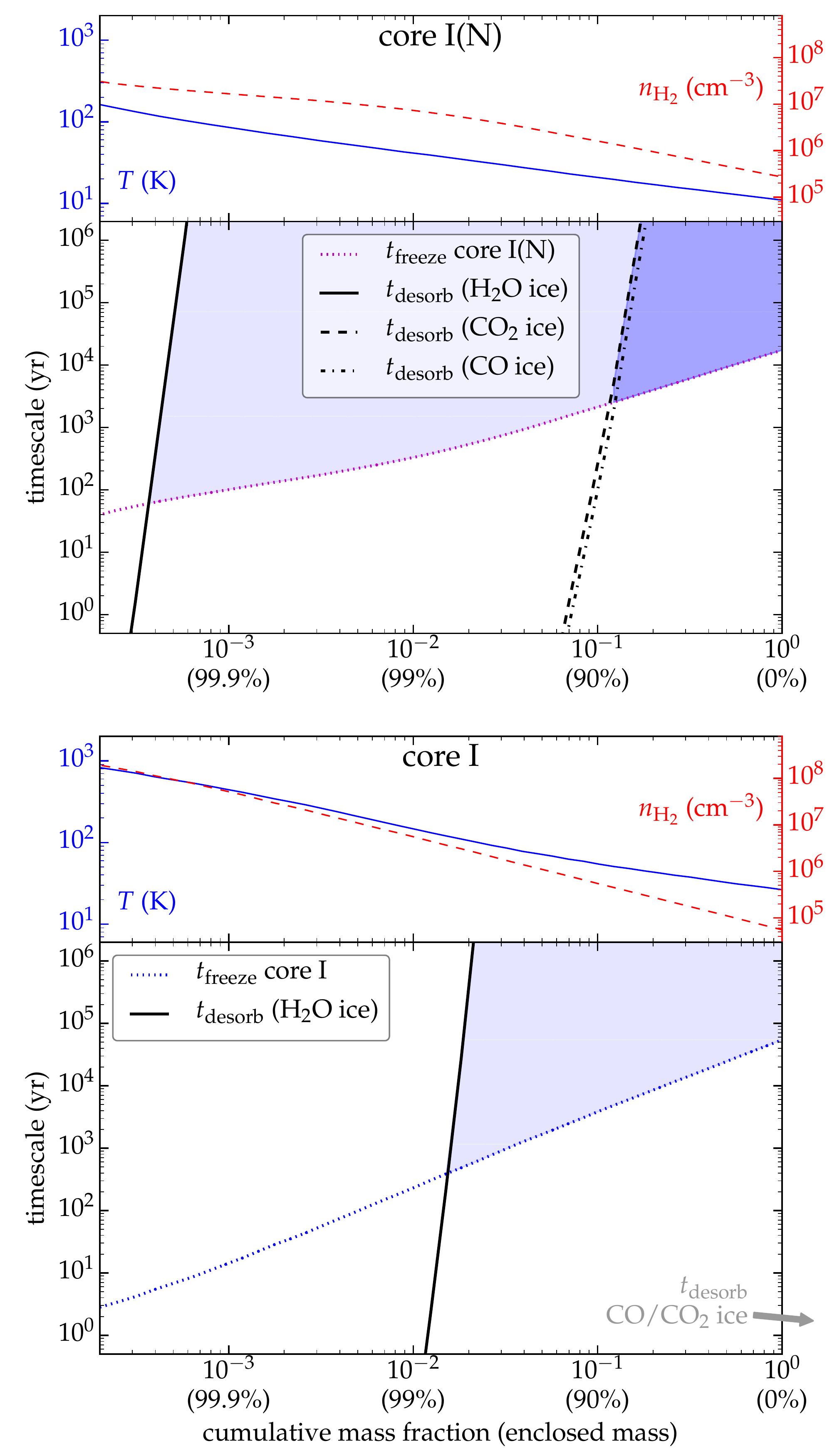}} 
	\caption{Radial dependence of adsorption and desorption timescales of HF from various types of grain surfaces. The top half of the figure relates to the envelope of core I(N), the bottom half to that of core I. The small panels above each main panel depict the physical structure (temperature in solid blue, left axis; density in dashed red, right axis) for the envelopes in question from \citet{vandertak2013a}. The horizontal axis is logarithmic cumulative mass increasing from left to right; labels in brackets indicate the percentage of enclosed mass starting from the outer shell of the core. Freeze-out of HF (depletion from the gas phase) occurs in the region where $t_\mathrm{desorb} > t_\mathrm{freeze}$; light shading indicates the case of \water\ ice mantles, darker shading that of CO/\COtwo\ ice mantles. The desorption time scale lines for CO ice and \COtwo\ ice in the bottom panel lie to the far right, outside of the limits of the axes. }
	\label{fig:HFphasediagram}
\end{figure}

In this work we consider multiple versions of the desorption timescale, because the binding energy in the exponent of Equation~\ref{eq:desorbrate} is heavily dependent on the type of grain surface. Unlike for more common molecular species such as CO \citep[e.g.,][]{bisschop2006,noble2012}, the desorption behavior of HF from astrophysically relevant grain surfaces has not been studied experimentally, so we rely on theoretical calculations. Typical interstellar dust grains, especially those embedded in cold, star-forming regions, are covered in one or multiple layers of ice consisting of various molecules, mainly \water, CO, and \COtwo\ \citep[for a recent review, see][]{boogert2015}. 
We collect binding energy values for several types of grain surfaces in Table~\ref{t:Ebinding}. For CO and \COtwo\ ice covered grains we adopt calculated binding energies from the literature \citep{chen2006,rivera-rivera2012}, while for hydrogenated bare silicate grains and \water\ ice covered grains, these values result from original ab initio chemical calculations performed for this work.

\begin{table}
\centering 
\caption{Binding energies for HF onto various surfaces.}
\label{t:Ebinding} 
\begin{tabular}{l l l l @{}}
\hline\hline
Type of grain surface							& $E_\mathrm{b,HF}$ & $E_\mathrm{b,HF}/k$	& Ref.\tablefootmark{a} \\
										& (kJ/mol)			& ($10^3$ K)	& \\
\hline  
Hydrogenated crystaline silica\tablefootmark{b}		& 9.2				& 1.1			& [1] \\  
\water\ ice on amorphous silica\tablefootmark{c}	& 53				& 6.3			& [1] \\   
CO ice on amorphous silica\tablefootmark{d}		& 8.9				& 1.07		& [2] \\   
\COtwo\ ice on amorphous silica\tablefootmark{d}	& 9.3				& 1.12		& [3] \\ 
\hline 
\end{tabular}
\tablefoot{
\tablefoottext{a}{References: [1] this work; [2] \citet{rivera-rivera2012}; [3]~\citet{chen2006}.} 
\tablefoottext{b}{For bonds with an SiH terminus.} 
\tablefoottext{c}{For bonds with an SiOH terminus.}
\tablefoottext{d}{Calculations for the binding energy of HF with CO and \COtwo\ were performed for the gas phase. See text for details of applicability to CO and \COtwo\ in solid form. } 
} 
\end{table}

To calculate the HF binding energies in the first two rows of Table~\ref{t:Ebinding} (with a bare grain and with \water\ ice), we carry out quantum calculations within the Kohn-Sham implementation of Density Functional Theory using the Quantum Espresso Simulation Package \citep{giannozzi2009}. Perdew-Burke-Ernzerhof exchange-correlation functional ultrasoft pseudopotentials are used. KS valence states are expanded in a plane-wave basis set with a cutoff at 340 eV for the kinetic energy.
The self-consistency of the electron density is obtained with the energy threshold set to $10^{-5}$ eV.  Calculations are performed using the primitive unit cell containing a total number of 46 atoms for bare hydrogenated silica, and 54 atoms for hydrogenated silica covered with one layer of \water\ ice. The geometry optimization is used within the conjugate gradients scheme, with a threshold of 0.01 eV\,$\AA^{-1}$ on the Hellmann-Feynman forces on all atoms; the Si atoms of the bottom layers are fixed at their bulk values. 
The binding energy of HF with the SiH terminus of hydrogenated crystalline silica is based on calculations for the hydroxylated alpha-quartz (001) surface. The binding energy of HF with one layer of \water\ ice on amorphous hydrogenated silicate is estimated by assuming that the the most common structure in this case is the HF molecule interacting with a \water\ molecule bonded to silanol (SiOH), which is the most abundant surface group in amorphous silica \citep{ewing2014}. 

The binding energies of HF with CO and \COtwo\ ice (last two rows of Table~\ref{t:Ebinding}) are taken from calculations by \citet{rivera-rivera2012} and \citet{chen2006}, respectively. These authors performed calculations for molecules in the gas phase. We consider the gas phase binding energies of HF with CO and \COtwo\ to be similar to those of HF with CO and \COtwo\ ices adsorbed on an inert surface such as that of hydroxylated amorphous silica. This approximation is based on the weak interactions of these ices with hydroxylated silica and within the CO and \COtwo\ molecular solids, so that the electronic density of CO and \COtwo\ in solid form is not significantly altered with respect to their state in the gas phase. Hence, for the aim of the present work, the binding energy of the HF molecule with CO or \COtwo\ as calculated in the gas phase is applicable for the condensed phase. 
The situation is notably different for interactions with \water\ in the gas or adsorbed form, because of its stronger interaction with silica and HF. For HF interacting with \water\ ice, we use binding energies from our own calculations described above. 

With the physical structure of both envelopes, the equations for the adsorption/desorption balance, and the binding energy values, the `freeze-out' region within each envelope is calculated in Fig.~\ref{fig:HFphasediagram}. Defining $t_\mathrm{freeze} = 1/\lambda$ and $t_\mathrm{desorb}=1/\xi$, HF molecules will deplete from the gas phase in the region of the envelope wherever $t_\mathrm{desorb}$ > $t_\mathrm{freeze}$. 
Numerical simulations by \citet{das2016} suggest that, within a mixed-composition ice layer, the abundances of CO and \COtwo\ are enhanced compared to \water\ in high-density ($\gtrsim10^5$~\pccm) environments. At such densities applicable for the protostellar envelopes studied here, we thus expect the HF binding energy to lie close to, but slightly above that of pure CO or \COtwo\ ice, and the true desorption time scale line in Fig.~\ref{fig:HFphasediagram} therefore somewhat to the left of the dash-dotted line for CO ice. In this case, HF is expected to stay frozen onto grain surfaces at a broad range of radii in core I(N): cumulative mass $\sim$0.1 to 1, i.e., 90\% of the mass, where the temperature is $\lesssim$20~K. 
Core I(N) is overall colder than core I both in the envelope (blue solid lines in Fig.~\ref{fig:HFphasediagram}) and in the embedded subcores \citep{hunter2006}.  For the comparatively warmer envelope of core I, the lines for the desorption timescale of HF from CO/\COtwo\ ice fall off the scale on the right hand side of the axes, leaving no freeze-out zone in this envelope. This could explain why HF is seen in the gas phase in core I, but not in core I(N). 

In an alternative scenario in which the dust grains are covered in pure \water\ ice --~or the desorption characteristics of a mixed mantle are dominated by that of \water\ ice \citep{collings2004}~-- the binding energy of HF with the ice would be greatly increased (see Table~\ref{t:Ebinding}). In this case, the HF freeze-out zone would expand to cover $>$98\% of the mass for \emph{both} envelopes, and our observations should have revealed no gas phase HF in either of the cores. Our observations are therefore inconsistent with pure \water\ ice coating on the grains. Instead, our interpretation relies on a significant part of the ice coating to consist of CO and/or \COtwo\ molecules. 
In principle, the composition of ice coatings do not need to be the same in the two neighboring envelopes. Particularly, given the lower temperatures of core I(N), there is higher probability that a significant amount of \water\ is frozen out in that envelope. This, again, would enhance the binding energy of HF onto the ice-covered dust grains in the envelope of core I(N), which may help to explain the lack of gas-phase HF observed toward source I(N). 

Conversely, if additional mechanisms for desorption, e.g., induced by (UV) photons or cosmic rays, would be taken into account, $\xi$ would increase, and the freeze-out region would be pushed to larger radii in the envelopes. Particularly UV photodesorption could have a different effect in one envelope compared to the other, because core I is more evolved and contains an \Hii\ region. 
An increased total desorption rate would have no effect on HF freeze-out in the envelope of core I, in which HF is already completely in the gas phase, but would reduce the size of the freeze-out zone for envelope I(N). If, however, thermal desorption as expressed in Equation~\ref{eq:desorbrate} is the dominant desorption mechanism and the binding energy of HF onto the ice surface is close to that of CO or \COtwo\ ice (Table~\ref{t:Ebinding}), our model predicts significant freeze-out of HF in core I(N), and none in core~I. 
It is important to note that, indeed, the temperature and density conditions under which HF remains frozen onto dust grains depend greatly on the exact composition and mixing of the ice mantle and therefore on the chemical history. It has already been recognized for example for the CO molecule that the freeze-out temperature `threshold' can vary considerably from one object to another \citep{qi2015b}.

\section{Conclusions}
\label{sec:conclusions}

In this work we present a map of HF absorption toward the northern end of the molecular cloud NGC\,6334, containing two well studied massive star-forming cores I and I(N). 
Although in the original definition of the observing program it was not anticipated that hydride absorption lines would be found within these data, the discovery space provided by the enormous frequency coverage of the {\it Herschel} SPIRE iFTS instrument has made this study possible. Such wide coverage in the far-infrared/submillimeter is only attainable with broadband FTS spectrometers \citep[see][for a review]{naylor2013}. 

The absorption line of HF is detected in 80\% of our mapped area, although it is spectrally unresolved by SPIRE. By complementing the new SPIRE iFTS data with existing, single pointing, high spectral resolution spectra from the {\it Herschel} HIFI instrument \citep{vanderwiel2010,emprechtinger2012}, we construct a three-dimensional picture of gas clouds in front of and inside the massive star-forming filament NGC\,6334. 

We find that our observations are consistent with a scenario of four individual foreground clouds on the line of sight toward NGC\,6334~I and I(N), two of which are unrelated to the star-forming complex (Sect.~\ref{sec:distinguishforeground}). The other two clouds are posited to be close to the dense molecular filament based on their spatial morphology. Their velocities are such that they are moving toward the star-forming cloud and could be adding several hundreds of solar masses of gas to the dense filament and the embedded cores in which massive star formation is already ongoing (Sect.~\ref{sec:foregroundrelation}). This component of gas is detected in rotational lines of diatomic hydride molecules, but had been unseen in studies of traditional dense gas tracers. In fact, using such a tracer, \HCOplus, \citet{zernickel2013} have inferred that the roughly cylindrically shaped NGC\,6334 filament is collapsing along its longest axis. Our work now indicates that accretion may also be ongoing in the perpendicular (radial) direction. Future studies of (competitively) accreting high-mass star-forming cores may need to take into account this additional low-density phase of the gas reservoir.  

Finally, in Sect.~\ref{sec:freezeout} we explain why HF is observed in the gas phase toward core I, but appears completely absent in core I(N). For this purpose, we use a simple description of adsorption and desorption time scales for HF interacting with dust grain surfaces, depending on the (radially variable) density and temperature. Since interactions of HF with interstellar-like dust grains have not been studied in the laboratory, we adopt binding energy values for different types of grain surfaces from theoretical calculations from the literature as well as from original work first presented in this paper. The conclusion is that the lower temperature of core I(N) compared to core I could lead to freeze-out of HF exclusively in the former, but only if the binding energy of HF onto the grain surface is governed by that of CO or \COtwo\ ice on a silicate surface. In this case, at the densities relevant in the envelope of source I(N) ($>$\pow{3}{5}~\pccm), we find that HF freezes out in the region of the envelope where the temperature is below $\sim$20~K, rather similar to the freeze-out temperature often adopted for CO. 
In contrast, if \water\ is the dominant constituent in the ice mantles, our model predicts that HF should have been frozen out at all radii in the envelopes of both sources I(N) and I. Since we observe a significant amount of HF in the gas phase in source I, this scenario is inconsistent with our data. 

Summarizing, this work uses HF as a sensitive tracer for (molecular) gas at relatively low densities that may be contributing mass to star forming cores. The HF signature reveals a gas reservoir that is inconspicuous in traditional dense gas tracers such as CO. In addition, we show that gas phase HF in higher density environments ($>$$10^5$~\pccm) is extremely sensitive to interactions with dust grains and will be depleted significantly at low dust temperatures.

\begin{acknowledgements} 
The research of MHDvdW at the University of Lethbridge was supported by the Canadian Space Agency (CSA) and the Natural Sciences and Engineering Research Council of Canada (NSERC), and at the University of Copenhagen by the Lundbeck Foundation. Research at the Centre for Star and Planet Formation is funded by the Danish National Research Foundation and the University of Copenhagen's programme of excellence. \\
SPIRE has been developed by a consortium of institutes led by Cardiff University (UK) and including Univ.~Lethbridge (Canada); NAOC (China); CEA, LAM (France); IFSI, Univ. Padua (Italy); IAC (Spain); Stockholm Observatory (Sweden); Imperial College London, RAL, UCL-MSSL, UKATC, Univ.~Sussex (UK); and Caltech, JPL, NHSC, Univ. Colorado (USA). This development has been supported by national funding agencies: CSA (Canada); NAOC (China); CEA, CNES, CNRS (France); ASI (Italy); MCINN (Spain); SNSB (Sweden); STFC, UKSA (UK); and NASA (USA). 
HIPE is a joint development by the Herschel Science Ground Segment Consortium, consisting of ESA, the NASA Herschel Science Center, and the HIFI, PACS and SPIRE consortia. \\
This research has made use of NASA's Astrophysics Data System Bibliographic Services. 
The graphical representations of the results in this paper were created using APLpy, an open-source plotting package for Python hosted at \mbox{\url{http://aplpy.github.com}}, Astropy, a community-developed core Python package for Astronomy (Astropy Collaboration \citeyear{astropy2013}), and the matplotlib plotting library \citep{matplotlib2007}. \\
The authors are grateful to Alexander Zernickel for providing and discussing excerpts of the CHESS spectral survey data, 
to Raquel Monje for providing the HF spectral model component profiles, 
and to Floris van der Tak for providing the physical structure models of the two envelopes in electronic table format.  
We thank Tommaso Grassi, Wing-Fai Thi, Jes J\o rgensen, S\o ren Frimann, and Mihkel Kama for discussions. 
\end{acknowledgements}

\bibliographystyle{aa}  
\bibliography{../../literature/allreferences}

\begin{appendix}

\section{Complementary figures}

\begin{figure} 
	\resizebox{\hsize}{!}{\includegraphics{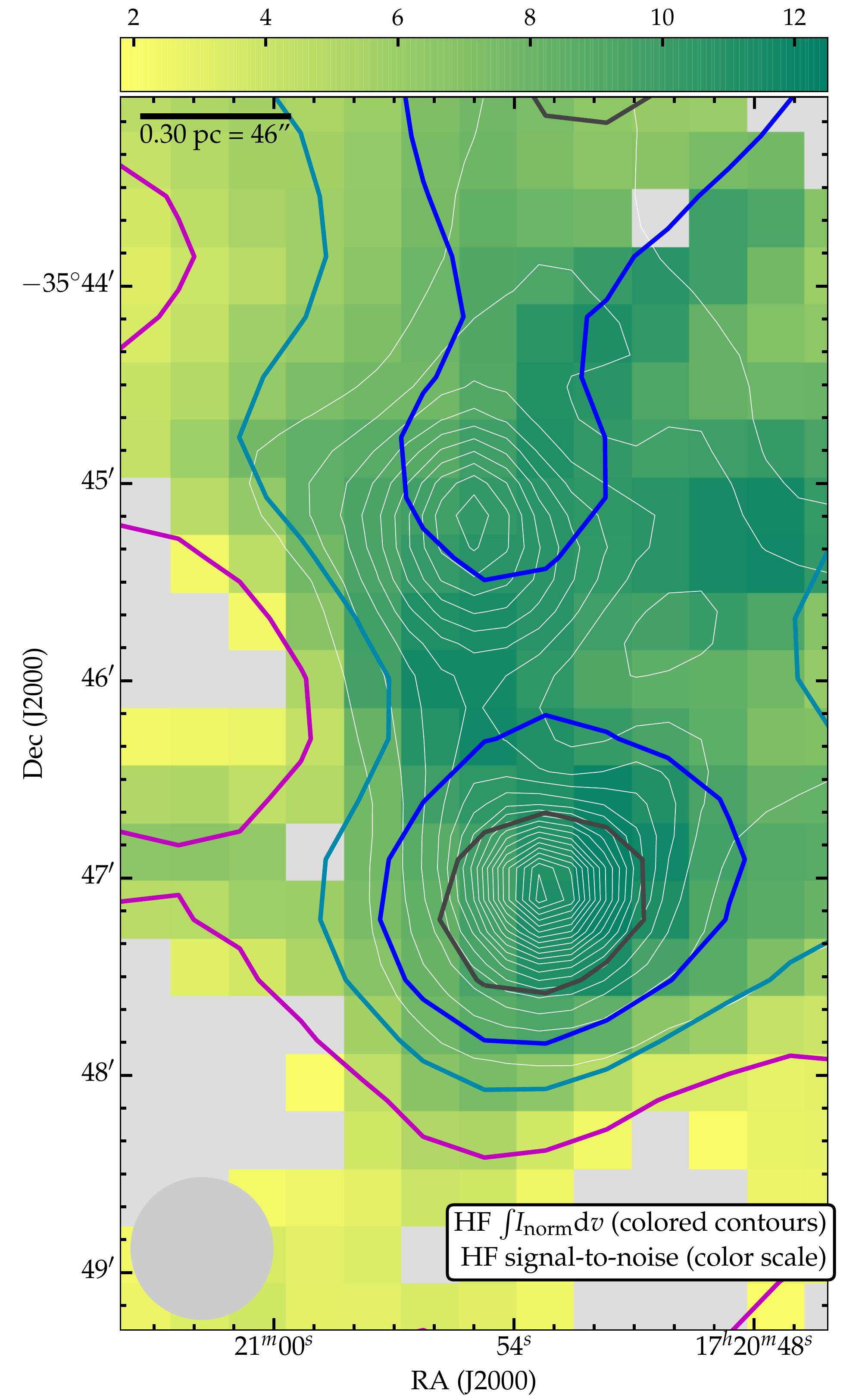}} 
	\caption{Signal-to-noise maps (color scale, with gray indicating values below 2) of the absorption signature of HF~1$\leftarrow$0. Colored contours are at the same equivalent width levels as in Fig.~\ref{fig:HFabsmap}. As in Fig.~\ref{fig:HFabsmap}, white contours trace continuum dust emission and the beam of the SPIRE iFTS map is shown in the bottom left corner. }
	\label{fig:SNmapHF}
\end{figure}

\begin{figure} 
	\resizebox{\hsize}{!}{\includegraphics{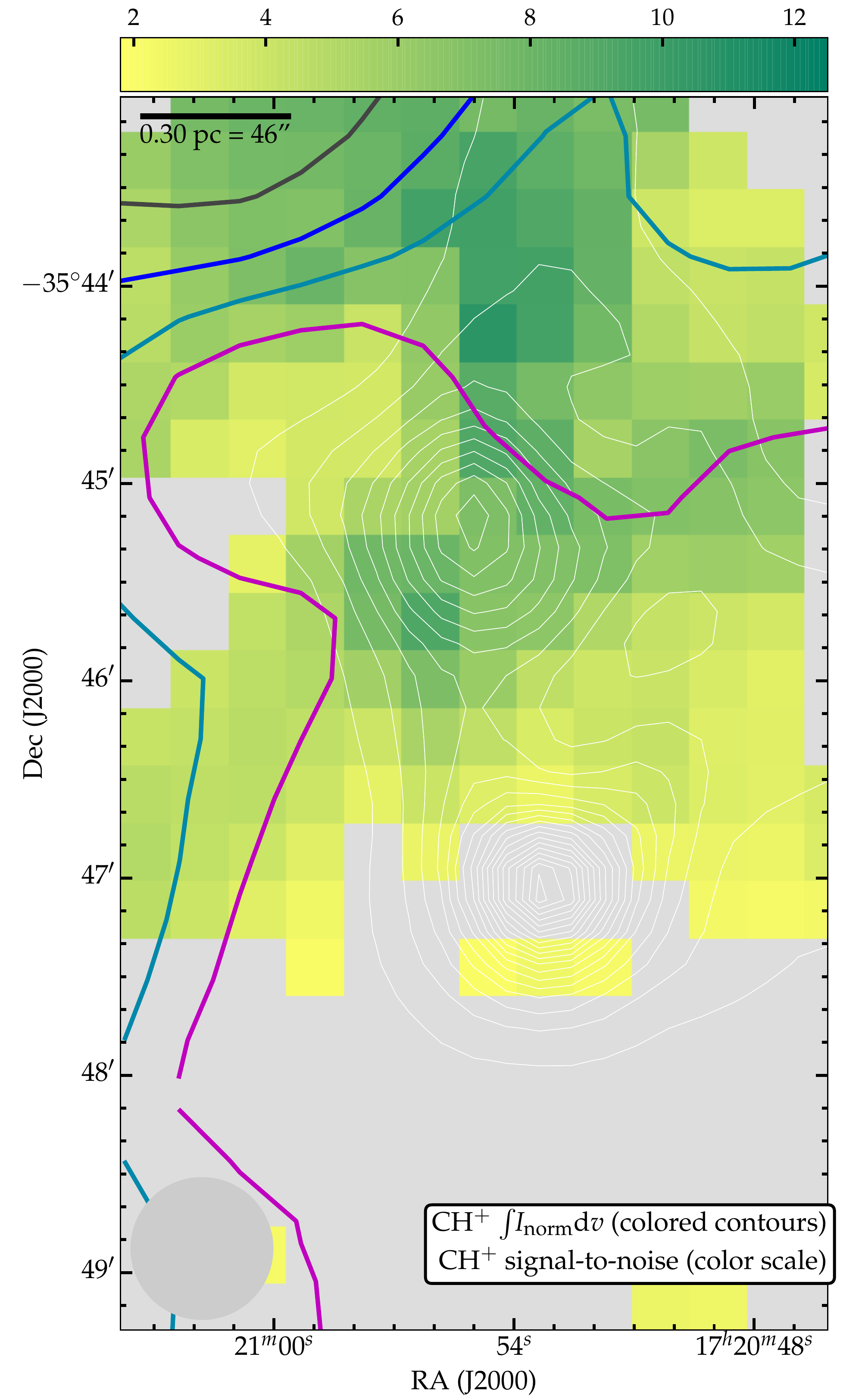}} 
	\caption{As in Fig.~\ref{fig:SNmapHF}, but for \CHplus~1$\leftarrow$0. Colored contours are at 22, 19, 16 and 13 \kms\ in gray, blue, cyan and magenta, respectively. }
	\label{fig:SNmapCHplus}
\end{figure}

Maps displaying the signal-to-noise ratio of the detections of the HF and \CHplus\ lines in each spatial pixel of our spectral cube (Sect.~\ref{sec:obsresults}) are shown in  Fig.~\ref{fig:SNmapHF} and \ref{fig:SNmapCHplus}. The colored contours in Fig.~\ref{fig:SNmapCHplus} show that \CHplus\ absorption is only confidently detected (signal-to-noise $>$ 5) in the northeastern section of the map. 

\begin{figure} 
	\resizebox{\hsize}{!}{\includegraphics{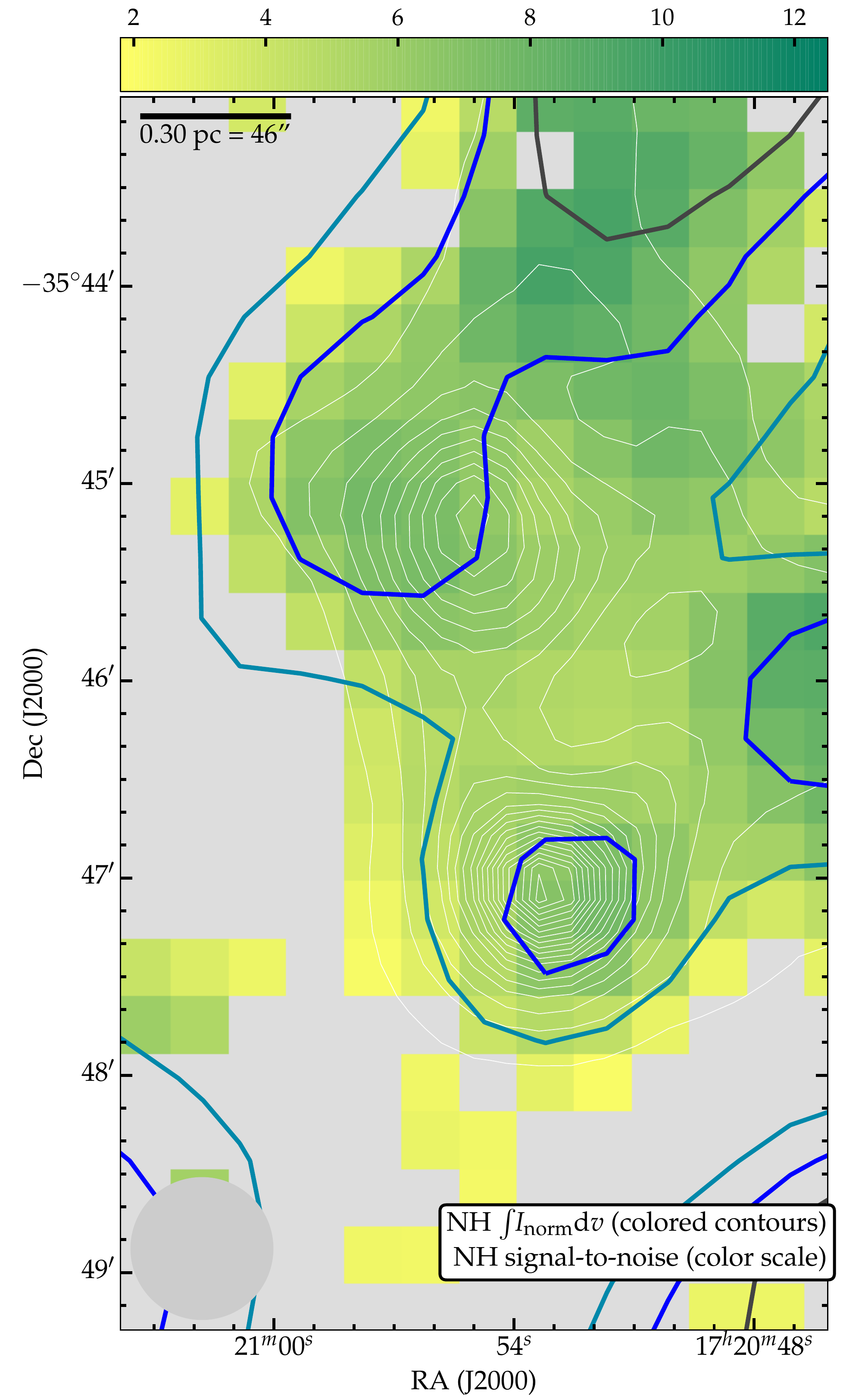}} 
	\caption{As in Fig.~\ref{fig:SNmapHF}, but for the $J$=2$\leftarrow$1 $N$=1$\leftarrow$0 transition of NH at 974.47~GHz. The $J$=1--1 transition at 1000.0~GHz shows a similar absorption distribution, but with detections in fewer pixels. Colored contours are at 16.5, 13.5, and 10.5~\kms\ in gray, blue, and cyan, respectively. }
	\label{fig:SNmapNH}
\end{figure}

\begin{figure} 
	\resizebox{\hsize}{!}{\includegraphics{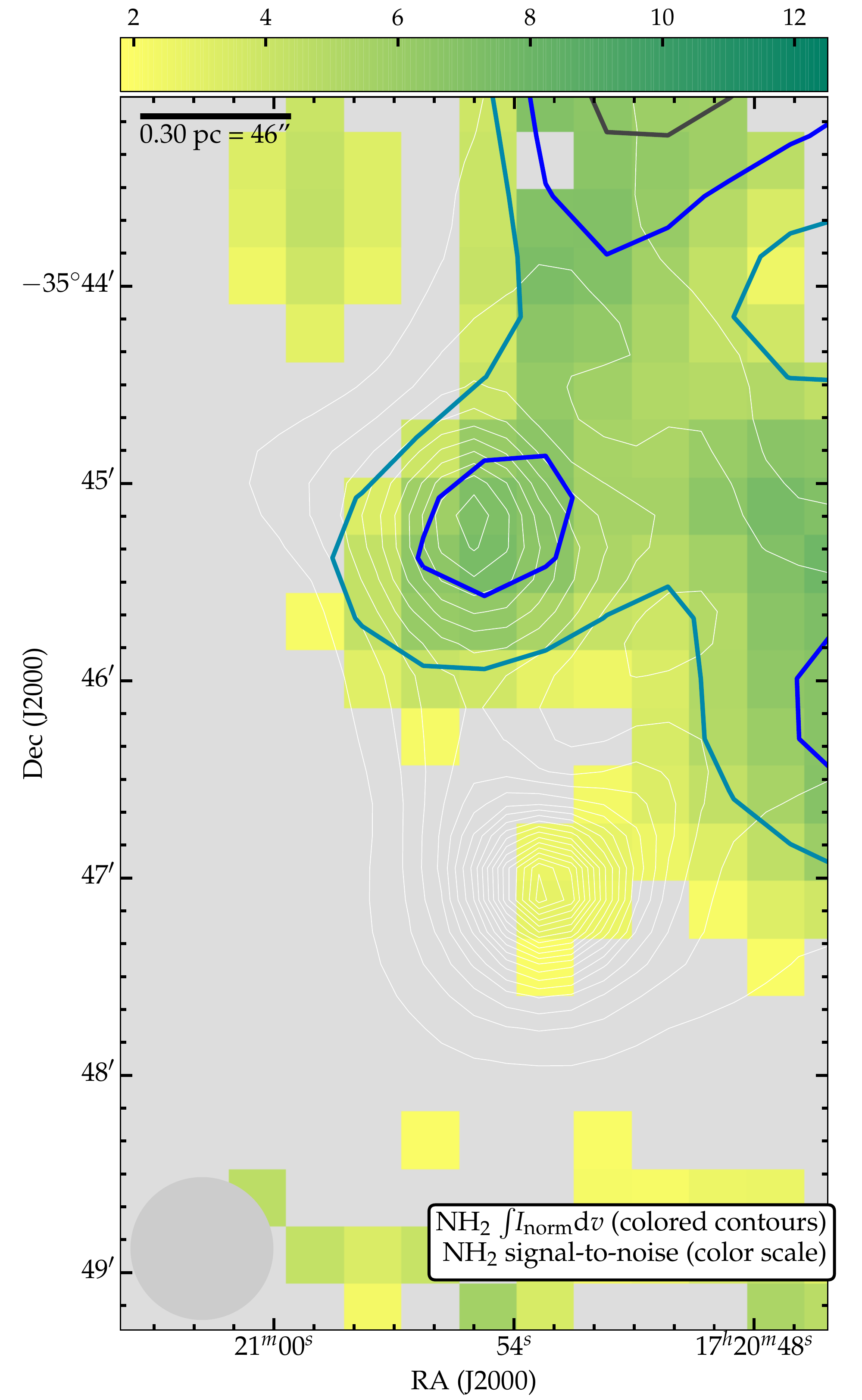}} 
	\caption{As in Fig.~\ref{fig:SNmapHF}, but for the $1_{11}$$\leftarrow$$0_{00}$~$J$=3/2$\leftarrow$1/2 transition of NH$_2$ at 952.57~GHz. The $J$=1/2$\leftarrow$1/2 transition at 959.50~GHz shows a similar absorption distribution, but with detections in fewer pixels. Colored contours are at 15.0, 11.5, and 8.0~\kms\ in gray, blue, and cyan, respectively. } 
	\label{fig:SNmapNH2}
\end{figure}

In addition, although the signal from nitrogen species is not interpreted in this paper, maps of line absorption depth due to NH and NH$_2$ are shown in Figs.~\ref{fig:SNmapNH} and \ref{fig:SNmapNH2}. The continuum-normalized cube for these lines is created -- analogous to those for HF and \CHplus~in Sect.~\ref{sec:obsresults} -- by fitting the continuum and the CO~9--8 line in the 945--1055~GHz section of the SSW cube. Absorption lines of NH (at 974.47 and 999.98~GHz) and two of NH$_2$ (at 952.57 and 959.50~GHz) are then fitted simultaneously with emission lines due to \thCO~9--8 at 991.3~GHz and \mbox{\water~$2_{02}$--$1_{11}$} at 987.9~GHz.

\begin{figure}[!hbt]
	\resizebox{\hsize}{!}{\includegraphics{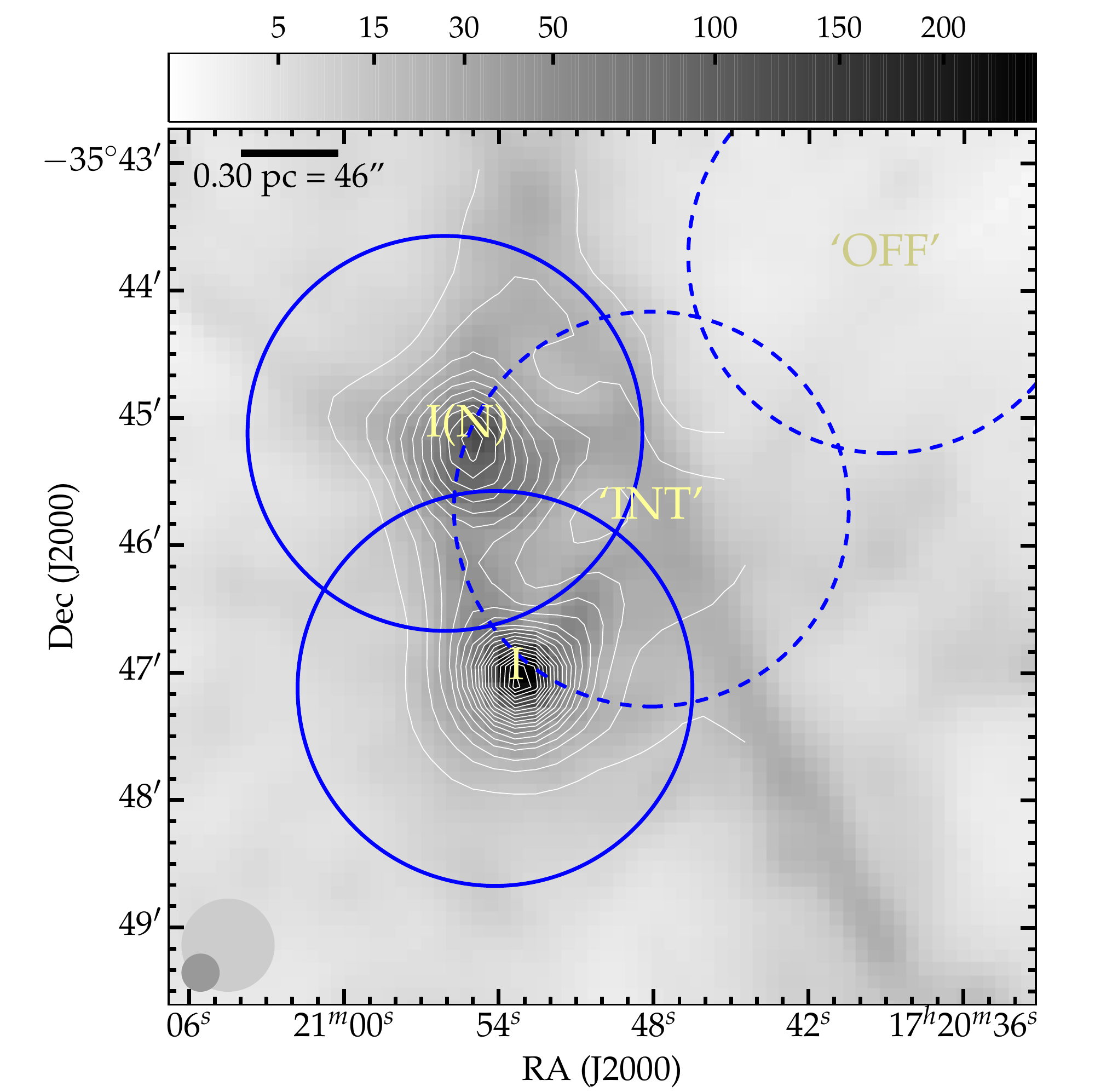}} 
	\caption{Placement of the SPIRE iFTS footprints on the NGC\,6334 region (cf.~Table~\ref{t:obs}). Solid circles indicate the fully sampled observations, two centered on cores I and one on I(N), while the sparsely sampled `INT' and `OFF' observations are marked by dashed circles. Circles are drawn with a diameter of 3\farcm1 to include the vignetted outer ring of the detector arrays \citep{spirehandbook2014}. The grayscale represents the same 250~\micron\ dust continuum map as in Fig.~\ref{fig:HFabsmap}, but with the scale bar stretching from 1 to 250  GJy\,sr$^{-1}$. The thin white contours, the beam size indicators in the bottom left and the scale bar in the top left corners are the same as in Fig.~\ref{fig:HFabsmap}. } 
	\label{fig:FTSfootprints}
\end{figure}

\end{appendix}

\end{document}